\documentclass[12pt, draftclsnofoot, onecolumn]{IEEEtran}
\ifCLASSINFOpdf
\else
\fi
\usepackage{graphicx}
\usepackage[numbers,sort&compress]{natbib}
\usepackage{amsthm}

\usepackage{amsmath}
\usepackage{amssymb}
\usepackage{graphicx}
\usepackage[space]{grffile}
\usepackage{subfig}

\theoremstyle{remark}
\newtheorem{thm}{$\;\;\;$Theorem}
\newtheorem{lem}{$\;\;\;$Lemma}
\newtheorem{prop}{$\;\;\;$Proposition}
\newtheorem{cor}{$\;\;\;$Corollary}
\newtheorem{defn}{$\;\;\;$Definition}

\begin{document}
%
\title{Signaling Design of Two-Way MIMO Full-Duplex Channel: Optimality Under Imperfect Transmit Front-End Chain}
%
%
%

\author{Shuqiao~Jia and
        Behnaam~Aazhang,~\IEEEmembership{}}

%
%

\markboth{draft}%
{Shell \MakeLowercase{\textit{et al.}}: Bare Demo of IEEEtran.cls for Journals}
%



\maketitle

\begin{abstract}
We derive the optimal signaling for a multiple input multiple output (MIMO) full-duplex two-way channel under the imperfect transmit front-end chain. We characterize the two-way rates of the channel by using a game-theoretical approach, where we focus on the Pareto boundary of the achievable rate region and Nash equilibia (NE). For a MISO full-duplex two-way channel, we prove that beamforming is an optimal transmission strategy which can achieve any point on the Pareto boundary. Furthermore, we present a closed-form expression for the optimal beamforming weights. In our numerical examples we quantify gains in the achievable rates of the proposed beamforming over the zero-forcing beamforming. For a general MIMO full-duplex channel, we establish the existence of NE and present a condition for the uniqueness of NE. We then propose an iterative water-filling algorithm which is capable of reaching NE. Through simulations the threshold of the self-interference level is found, below which the full-duplex NE outperforms the half-duplex TDMA.

\end{abstract}


%
\IEEEpeerreviewmaketitle

\section{Introduction}
A node in a full-duplex mode can simultaneously transmit and receive in the same frequency band. Therefore, the wireless channel between two full-duplex nodes can be bidirectional, having the potential to double the spectral efficiency when compared to the half-duplex network. Due to the proximity of the transmit and receive antennas on a node, the overwhelming self-interference becomes the fundamental challenge in implementing a full-duplex network. The mitigation of the self-interference signal can be managed at each step of the communication network by passive and active cancellation methods \cite{sahai2013ITVT}. In recent work \cite{Everett2011Allerton,Everett2011Asilomar,bharadia2013sigcomm}, the feasibility of the single input single output (SISO) full-duplex communication has been experimentally demonstrated. However, the performance is limited by the residual self-interference which is considered in \cite{sahai2013ITVT,bharadia2013sigcomm,vehkapera2013PIMRC,Day2012ITSP} to be induced by the imperfection of the transmit front-end chain.

The performance bottleneck from imperfect transmit front-end chain has motivated recent research in full-duplex channel with transmit front-end noise. The performance of the SISO full-duplex two-way channel has been thoroughly analyzed in \cite{Duarte2012ITWC,sahai2013ITVT}. The multiple input multiple output (MIMO) full-duplex two-way channel with transmit front-end noise is considered in \cite{vehkapera2013PIMRC,Day2012ITSP,Cirik2015weighted} (in \cite{vehkapera2013PIMRC,Day2012ITSP} termed as MIMO full-duplex bidirectional channel). In \cite{vehkapera2013PIMRC}, the transmit front-end noise was modeled as a white Gaussian noise. Following the transmission noise model, the effect of time-domain cancellation and spatial-domain suppression on a full-duplex channel were studied. In \cite{Day2012ITSP}, a full-duplex channel was modeled with the transmit front-end noise and under the limited dynamic range. The authors then proposed a numerical method to solve the signaling that maximizes the lower bound of achievable sum-rate for such a full-duplex channel. The maximization of the weighted achievable sum-rate for a full-duplex channel was considered under the imperfect transmit front-end chain in \cite{Cirik2015weighted}.

Within this context, we consider optimally operating a full-duplex channel under imperfect front-end chains. We introduce a full-duplex channel model that includes the effect of imperfect transmit front-end chain and limited transmitter dynamic range. Such a channel model is closely related to a Gaussian interference channel model, which were widely studied in \cite{scutari2008competitive,scutari2009mimo,palomar2010convex}. Inspired by the work in \cite{scutari2008competitive,scutari2009mimo,palomar2010convex}, we consider a full-duplex two-way channel in a game-theoretical framework. Consequently, we characterize a full-duplex channel by Pareto boundary and Nash equilibrium. In game theory, Pareto boundary is a definition with the global optimality whereas Nash equilibrium is with the competitive optimality. Unlike the global optimality, competitive optimality is a definition of optimality that can be achieved by distributed algorithms. 

For a MIMO full-duplex channel, the Pareto boundary of the achievable rate region is described by a family of non-convex optimization problems. In the special case, where there is only a single receive antenna, we can decouple the original non-convex problems to a family of convex optimization problems \cite{Shang2011ITInf,boyd2009convex}. By employing the semi-definite programing (SDP) reformulation, we then numerically solve the Pareto-optimal signaling by which the Pareto boundary of a MISO full-duplex channel can be achieved. We further prove that the rank of Pareto-optimal signaling is always equal to one. That is to say, for a MISO full-duplex two-way channel, transmit beamforming scheme is capable of achieving the entire Pareto boundary. Furthermore, we propose a closed-form for the optimal beamforming weights. 

The Pareto boundary of a general MIMO full-duplex channel cannot be decoupled or transformed into a convex form. It implies that, to find the Pareto boundary, a family of centralized nonconvex problems needs to be solved, which renders the computation intractable. Therefore, for a general MIMO full-duplex channel we restrict our attention to the optimality which can be achieved by fully distributed algorithms. In other words, instead of the Pareto boundary, we aim to achieve the Nash equilibrium for a general MIMO full-duplex channel. In this paper, we first prove the existence of the Nash equilibrium for a MIMO full-duplex channel. We then derive a condition to ensure the uniqueness of Nash equilibrium.
The signaling at the Nash equilibrium can be derived by our proposed algorithm, which is modified from the iterative water-filling algorithm (IWFA) in \cite{palomar2010convex}.


The rest of the paper is organized as follows. In Section \uppercase\expandafter{\romannumeral2} the channel model for a MIMO two-way full-duplex wireless channel is presented. Section \uppercase\expandafter{\romannumeral3} presents the description of the Pareto optimality and the competitive optimality, which correspond to the Pareto boundary and the Nash equilibrium, respectively. The characterizations of the Pareto boundary and the Nash equilibrium for a full-duplex channel are also provided. In Section \uppercase\expandafter{\romannumeral4} the Pareto boundary of a MISO full-duplex channel is derived, where the beamforming scheme is proved to be opitmal. The closed-form solution for the optimal beamforming weights is then presented. Section \uppercase\expandafter{\romannumeral5} presents the existence of the Nash equilibrium for a full-duplex channel. The condition for the uniqueness of NE is also provided. Here, we propose a modified iterative water-filling algorithm to achieve the NE. Numerical examples are provided in Section \uppercase\expandafter{\romannumeral6}. While the conclusions are given in Section \uppercase\expandafter{\romannumeral7}. 

\textit{Notation}: We use $(\cdot)^{\dag}$ to denote conjugate transpose. For a scalar $a$, we use $|a|$ to denote the absolute value of $a$. For a vector $\boldsymbol{a}\in \mathbb{C}^{M\times 1}$, we use $\|\boldsymbol{a}\|$ to denote the norm, $\boldsymbol{a}^{(k)}$ to denote the $k^{th}$ element of $\boldsymbol{a}$, $\text{Diag}(\boldsymbol{a})$ to denote the square diagonal matrix with the elements of vector $\boldsymbol{a}$ on the main diagonal. For a matrix $\boldsymbol{A}\in \mathbb{C}^{M\times M}$, we use $\boldsymbol{A}^{-1}$, $\textbf{tr}(\boldsymbol{A})$ and $\text{rank}(\boldsymbol{A})$ to denote  the inverse, the trace and the rank of $\boldsymbol{A}$, respectively. We use $\text{diag}(\boldsymbol{A})$ to denote the diagonal matrix with the same diagonal elements as $\boldsymbol{A}$. $\boldsymbol{A}\succeq 0$ means that $\boldsymbol{A}$ is a positive semidefinite Hermitian matrix. We denote expectation, variance and covariance by $\text{E}\{\cdot\}$, $\text{Var}\{\cdot\}$ and $\text{Cov}\{\cdot\}$, respectively. Finally, $\mathbb{C}$ and $\mathbb{H}$ denotes the complex field and the Hermitian symmetric space, respectively.

%
%
%
%

 \section{Channel Model}
 
 \begin{figure}[!t] 
 \centering
 \includegraphics[width=4in]{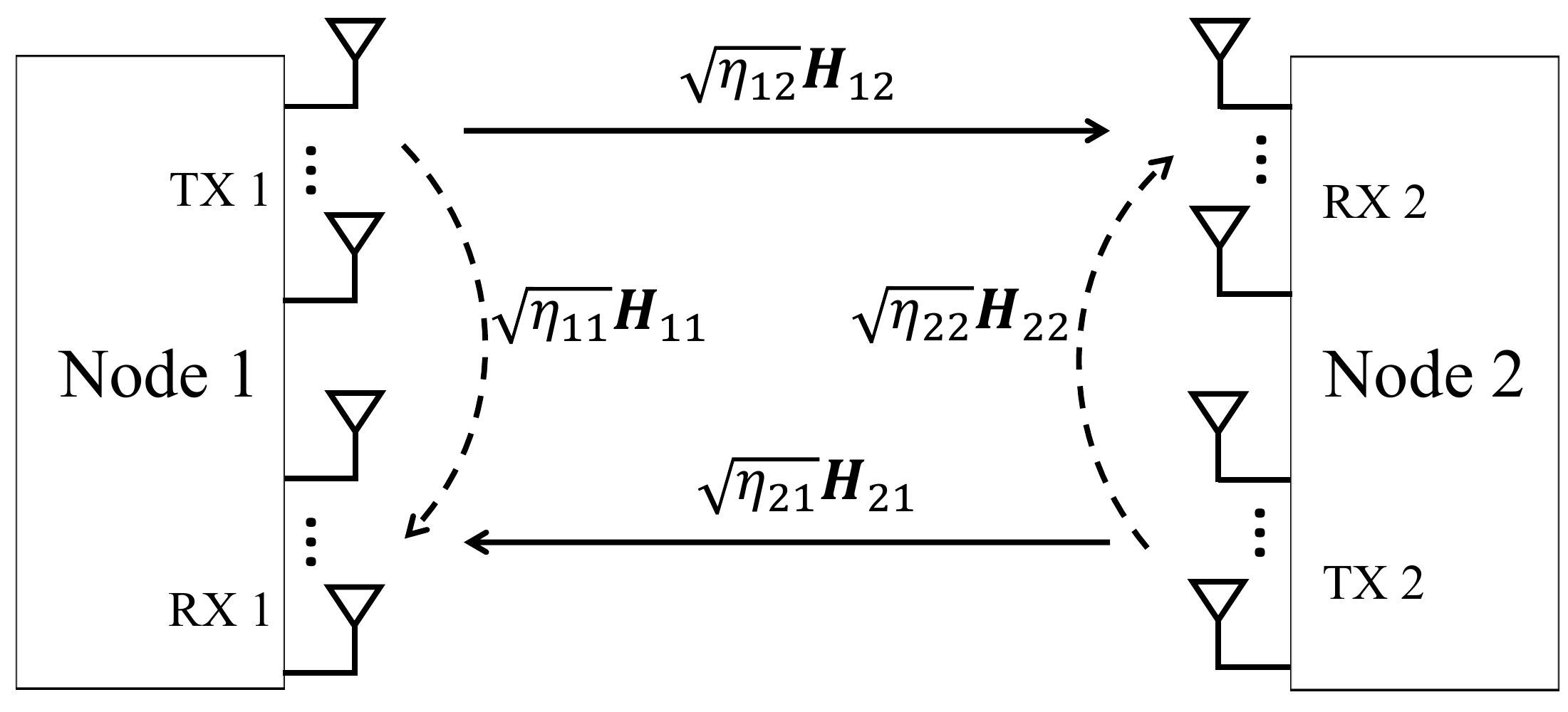}
 \caption{The MIMO point-to-point full-duplex network under study. The solid lines denote the direct channels and the dashed lines denote the self-interference channels.} 
 \label{Fig1}
 \vspace{-10pt}
 \end{figure}
 
 In this section, we present the channel model for a MIMO full-duplex (FD) network with two nodes as illustrated in Fig. \ref{Fig1}. We assume that two nodes indexed by $i,j\in\{1,2\}$ share the same single frequency band for transmission. Each node has a transmitter and a receiver. The transmitter is equipped with $M$ physical antennas and the receiver with $N$ physical antennas, where each single antenna is connected to a front-end chain. The signal from transmitter $i$ is collected as the signal of interest by receiver $j, j\neq i$, while appears at its own receiver $i$ as the self-interference signal.
 
 As illustrated in Fig. \ref{Fig1} the direct channel between two nodes is denoted by $\sqrt{\eta_{ij}}\boldsymbol{H}_{ij}, i\neq j$, where  $\eta_{ij}$ represents the average power gain of the direct channel. Similarly, the self-interference channel within each node is characterized by the channel matrix $\boldsymbol{H}_{ii}$ and the average power gain $\eta_{ii}$. According to \cite{Day2012ITSP}, all the channels in the above full-duplex network can be modeled as the Raleigh fading channel. That is, all channel matrices are with i.i.d complex Gaussian entries with zero mean and unit variance. We define $\gamma_{i}\triangleq\frac{\eta_{ji}}{\eta_{ii}}$ to represent the relative strength of the direct channel and the self-interference channel.
  
 While passing through the transmit front-end chain, the intended transmit signal is corrupted by distortions in the power amplifier, non-linearities in the DAC and phase noise \cite{bharadia2013sigcomm,vehkapera2013PIMRC}. The results in \cite{bharadia2013sigcomm,vehkapera2013PIMRC} demonstrate that all the impairments induced by the imperfect front-end chain can be comprehensively modeled by an additive Gaussian noise, namely, the transmit front-end noise. Furthermore, the power of the transmit frond-end noise is $\beta$ times proportional to that of the intended transmit signal due to the limited dynamic range of the transmit front-end chain \cite{bharadia2013sigcomm,Day2012ITSP}. Here, $\beta$ denotes the noise level of the transmit front-end chain \cite{Day2012ITSP}.
 
 \begin{figure}[!t] 
 \centering
 \includegraphics[width=4in]{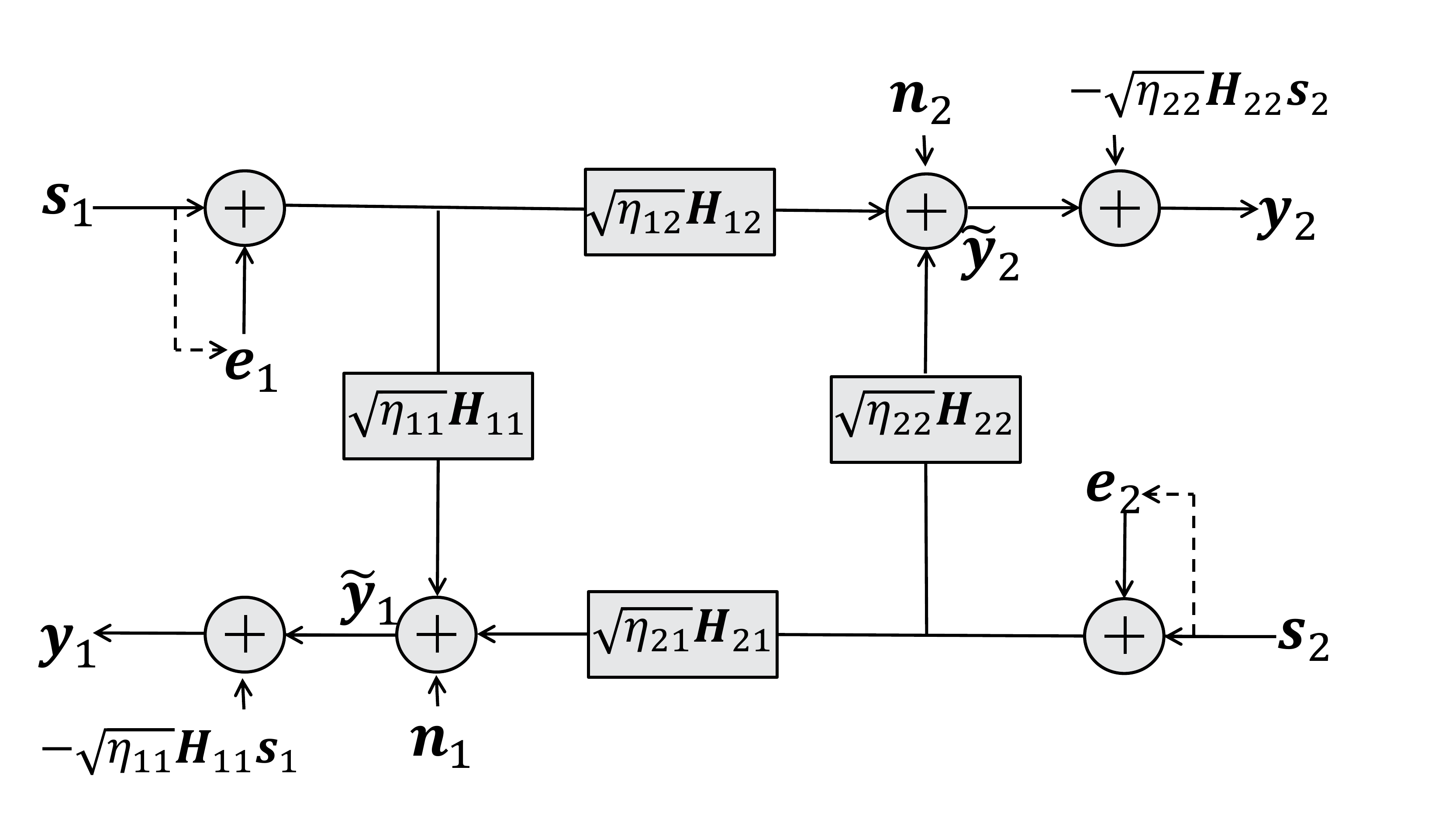}
 \caption{The MIMO point-to-point full-duplex network under study. The solid line denotes the desired channel and the dashed line denotes the self-interference channel.} 
 \label{Fig2}
 \vspace{-10pt}
 \end{figure}
 
 Fig. \ref{Fig2} summarizes our full-duplex channel model. The signal at receiver $i$ is given by
 \begin{equation}
    \widetilde{\boldsymbol{y}}_i=\sqrt{\eta_{ji}}\boldsymbol{H}_{ji}(\boldsymbol{s}_j+\boldsymbol{e}_j)+\sqrt{\eta_{ii}}\boldsymbol{H}_{ii}(\boldsymbol{s}_i+\boldsymbol{e}_i)+\boldsymbol{n}_i,
    \label{mimo received signal}
 \end{equation}
where $\boldsymbol{s}_i\in \mathbb{C}^{M\times1}$ denotes the intended transmit signal at transmitter $i$, the channel matrices $\boldsymbol{H}_{ij}\in\mathbb{C}^{N\times M}$. The transmit front-end noise $\boldsymbol{e}_i$ is propagated over the same channel as $\boldsymbol{s}_i$.  Denote the covariance of $\boldsymbol{s}_i$ by $\boldsymbol{Q}_i\triangleq\text{Cov}\{\boldsymbol{s}_i\}$. Note that the $m^{th}$ diagonal element of $\boldsymbol{Q}_i$ represents the transmit signal power of the $m^{th}$ antenna at transmitter $i$. Thus, $\boldsymbol{e}_i$ can be modeled as the Gaussian vector with zero mean and covariance $\text{Cov}\{\boldsymbol{e}_i\}=\beta \text{diag}(\boldsymbol{Q}_i)$ \cite{Day2012ITSP,Cirik2015weighted}. The thermal noise at receiver 
$i$ is modeled as $\boldsymbol{n}_i\sim\mathcal{CN}(0,\boldsymbol{I}_N)$, where $\boldsymbol{I}_N$ denotes the $N\times N$ identity matrix.

At receiver $i$, the signal of interest $\boldsymbol{H}_{ji}\boldsymbol{s}_{j}, j\neq i$ is received along with the self-interference signal $\boldsymbol{H}_{ii}\boldsymbol{s}_{i}$ and the transmit front-end noise $\boldsymbol{H}_{ji}\boldsymbol{e}_{j}$, $\boldsymbol{H}_{ii}\boldsymbol{e}_{i}$. The power level of $\boldsymbol{H}_{ji}\boldsymbol{e}_{j}, j\neq i$ is typically much lower than that of the thermal noise $\boldsymbol{n}_i$ and thus can be neglected \cite{vehkapera2013PIMRC}. However, $\boldsymbol{H}_{ii}\boldsymbol{e}_{i}$ is in the power level close to the signal of interest and needs to be considered for analysis, since the power gain of the self-interference channel $\boldsymbol{H}_{ii}$ overwhelms the power gain of the direct channel $\boldsymbol{H}_{ji}$ \cite{Day2012ITSP}. 

In addition to the strength, transmitters and receivers on a same node are relatively static, resulting in the long coherence time of self-interference channels, thus receiver $i$ is assumed to have the perfect knowledge of its own self-interference channel $\boldsymbol{H}_{ii}$ \cite{vehkapera2013PIMRC}. Note that receiver $i$ also knows its own transmitted signal $\boldsymbol{s}_i$. Then we can eliminate the self-interference $\boldsymbol{H}_{ii}\boldsymbol{s}_i$ before decoding. The signal after cancellation is given by
 
 \begin{equation}
     \boldsymbol{y}_i=\sqrt{\eta_{ji}}\boldsymbol{H}_{ji}\boldsymbol{s}_j+\sqrt{\eta_{ii}}\boldsymbol{H}_{ii}\boldsymbol{e}_i+\boldsymbol{n}_i,
     \label{equivalent channel model}
  \end{equation}
where $\boldsymbol{H}_{ii}\boldsymbol{e}_i$ represents the residual self-interference.

\section{Pareto Optimality and Competitive Optimality}
 As shown in (\ref{equivalent channel model}), the transmission from node $j$ to node $i$ is corrupted by the residual self-interference of node $i$ and the thermal noise. The sum of all such interferences is equal to an additive Gaussian noise $\boldsymbol{v}_i$. Let us define $\boldsymbol{\Sigma}_i$ to be the covariance of $\boldsymbol{v}_{i}$, then $\boldsymbol{\Sigma}_i=\boldsymbol{I}+\beta\eta_{ii}\boldsymbol{H}_{ii}\text{diag}(\boldsymbol{Q}_i)\boldsymbol{H}_{ii}^{\dag}$, which is a function of $\boldsymbol{Q}_i$. It follows from the results of \cite{telatar1999capacity,cover2012elements} that by employing a Gaussian codebook at node $1$, we can achieve the maximum rate for the transmission from node $1$ to node $2$
\begin{equation}
   R_1(\boldsymbol{Q}_1,\boldsymbol{Q}_2)=\text{log}\;\text{det}(\boldsymbol{I}+\eta_{12}\boldsymbol{H}_{12}^{\dag}\boldsymbol{\Sigma}_{2}^{-1}\boldsymbol{H}_{12}\boldsymbol{Q}_{1}),
   \label{R1}
 \end{equation}
where $(\boldsymbol{Q}_{1},\boldsymbol{Q}_2)$ are the transmit covariance matrices of the nodes. Similarly, the maximum rate for the transmission from node $2$ to node $1$ is equal to
\begin{equation}
R_2(\boldsymbol{Q}_1,\boldsymbol{Q}_2) =  
 \text{log}\;\text{det}(\boldsymbol{I}+\eta_{21}\boldsymbol{H}_{21}^{\dag}\boldsymbol{\Sigma}_{1}^{-1}\boldsymbol{H}_{21}\boldsymbol{Q}_{2}).
 \label{R2}
\end{equation}

Denote the feasible set of the covariance matrix $\boldsymbol{Q}_i$ as $\mathcal{X}_i$. A set of $(\boldsymbol{Q}_1,\boldsymbol{Q}_2)$ is feasible if it satisfies the transmit power constraints $P_1,P_2$. Thus, we have $\mathcal{X}_i=\left\lbrace\boldsymbol{Q}_i\in\mathbb{H}^{M}|\boldsymbol{Q}_i\succeq0,\textbf{tr}\left(\boldsymbol{Q}_i\right)\leq P_i\right\rbrace$. Once $(\boldsymbol{Q}_{1},\boldsymbol{Q}_2)$ are given, only rate pair $(r_1,r_2)$ with $r_1\leq R_1, r_2 \leq R_2$ is achievable for the FD channel. Thus, the achievable rate region for the MIMO FD two-way channel with the transmit power constraints $P_1,P_2$ can be described as the following set: 
\begin{equation}
\mathcal{R}\triangleq
\bigcup_{
\scriptstyle \;\boldsymbol{Q}_1\in\mathcal{X}_1,
\atop
\scriptstyle \boldsymbol{Q}_2\in\mathcal{X}_2}
\left\lbrace 
  \begin{split}
  &(r_1,r_2):\\
  & 0\leq r_1 \leq R_1(\boldsymbol{Q}_1,\boldsymbol{Q}_2)\\
  & 0\leq r_2 \leq R_2(\boldsymbol{Q}_1,\boldsymbol{Q}_2)
  \end{split}
  \right\rbrace,
\label{centralized region} 
\end{equation}
where $R_1$ and $R_2$ in (\ref{R1}) and (\ref{R2}), respectively, are mutually coupled by the covariance matrices $\boldsymbol{Q}_1$ and $\boldsymbol{Q}_2$. Therefore, there always exists performance tradeoffs between $R_1$ and $R_2$ in a selection of $(\boldsymbol{Q}_1, \boldsymbol{Q}_2)$. Such tradeoffs can be considered as a game in which node $i$ is player $i$, $R_i(\boldsymbol{Q}_1,\boldsymbol{Q}_2)$ is the payoff of player $i$ and $\boldsymbol{Q}_i$ is the admissible strategy of player $i$. All possible outcomes of the game are characterized in the achievable rate region $\mathcal{R}$. As a sequence, all concepts of a FD channel can be interpreted from a game-theoretic view. Driven by the global optimality, we first consider the Pareto boundary for a FD two-way channel. The Pareto boundary is characterized by a set of 'jointly' optimal rate pairs $(R_1,R_2)$. Each jointly optimal rate pair is of the Pareto-optimality, which is defined as follows (A similar definition can be found in \cite{Mochaourab2011ITSP,jorswieck2008ITSP,zhang2010ITSP}).

\begin{defn}[Pareto optimality]
A rate pair $(R_1^*,R_2^*)\in \mathcal{R} $ is Pareto optimal
if there does not exist another rate pair $(R_1,R_2)\in \mathcal{R}$ such that $(R_1,R_2)\geq (R_1^*,R_2^*)$ and $(R_1,R_2)\neq (R_1^*,R_2^*)$ where the inequality is component-wise.
\label{Definition Pareto optimality}
\end{defn}
The Pareto boundary refers to the outer boundary of the achievable rate region $\mathcal{R}$ in (\ref{centralized region}). Thus, we can define the Pareto boundary $\mathcal{R}^*$ as follows
\begin{equation}
\mathcal{R}^*=\bigcup \;\{\text{all the Pareto optimal rate pairs}\;\; (R_1^*, R_2^*)\;\;\text{in}\;\;\mathcal{R}\}.
\label{Pareto boundary}
\end{equation}
Each point on the Pareto boundary $\mathcal{R}^*$ maximizes one of the weighted sum-rates for the FD channel \cite{Shang2007Asilomar,Shang2011ITInf}. Therefore, $\mathcal{R}^*$ can be derived by solving a family of weighted sum-rate optimization problems:
\begin{equation}
\begin{split}
\max\limits_{\boldsymbol{Q}_1,\boldsymbol{Q}_2}\;\;& {\mu_1R_1(\boldsymbol{Q}_1,\boldsymbol{Q}_2)+\mu_2R_2(\boldsymbol{Q}_1,\boldsymbol{Q}_2)}\\
\text{subject to}\;\;
& \boldsymbol{Q}_i\in\mathcal{X}_i, i=1,2,
\end{split}
\label{non-convex boundary problems}
\end{equation}
where $0\leq\mu_1,\mu_2$ and $\mu_1+\mu_2=1$. 

The optimal solutions $(\boldsymbol{Q}_1^*,\boldsymbol{Q}_2^*)$ for problem (\ref{non-convex boundary problems}) with some $\mu_1,\mu_2$ correspond to one pair of Pareto-optimal transmission strategies for the FD channel. To obtain the entire Pareto boundary, we need to derive all Pareto-optimal strategies. However, the centralized non-convex nature of problem (\ref{non-convex boundary problems}) poses two serious issues in achieving the Pareto boundary. First, solving the Pareto-optimal strategies in general comes at the price of prohibitively high computationally complexity due to the non-convexity of problem (\ref{non-convex boundary problems}). Second, problem (\ref{non-convex boundary problems}) is coupled by $\boldsymbol{Q}_1,\boldsymbol{Q}_2$. Hence, it requires an extra central node to acquire the full knowledge of the FD channel and then solve the Pareto-optimal strategies. Due to these challenges, it is often not practical to operate a FD channel in its Pareto optimality. An alternative way is that each node would compete for its own payoff irrespective of the other node's payoff. The optimality built in such a scenario is defined as follows.

\begin{defn}[competitive optimality]
 For a two-way FD channel, $\mathcal{X}_i$ is the nonempty set of all feasible strategies for node $i$. A strategy profile $(\boldsymbol{Q}_1^*,\boldsymbol{Q}_2^*)\in \mathcal{X}_1\times \mathcal{X}_2$ is competitive optimal if the following condition holds for all $i,j\in\{1,2\}, i\neq j$:
\begin{equation}
R_i(\boldsymbol{Q}_i^*,\boldsymbol{Q}_j^*)\geq R_i(\boldsymbol{Q}_i, \boldsymbol{Q}_j^*), \forall\boldsymbol{Q}_i\in \mathcal{X}_i.
\label{competitive optimality}
\end{equation}
\label{Definition competitive optimality} 
\end{defn}
\vspace*{-20pt}
If the competitive optimality is achieved, any unilateral change of strategies would result in a rate loss for the FD channel \cite{palomar2010convex}. From the game theoretic view, a set of competitive optimal strategies corresponds to a Nash equilibrium (NE) of the FD channel. To obtain an NE, we construct a non-cooperative game according to Definition \ref{Definition competitive optimality}. In the game, node $i$ is assumed to have the knowledge of the direct channel $\boldsymbol{H}_{ij}$ and its own self-interference channel $\boldsymbol{H}_{ii}$, and have a fixed power budget $P_i$. At each iteration, given the strategy of node $j$, node $i$ locally chooses its strategy $\boldsymbol{Q}_i$ to maximize its pay-off $R_i$, which can be described by the rate-maximization problem as follows:
\begin{equation}
\begin{split}
\max_{\boldsymbol{Q}_i}\;\;& {R_i(\boldsymbol{Q}_i,\boldsymbol{Q}_j)}\\
\text{subject to}\;\;
& \boldsymbol{Q}_i\in\mathcal{X}_i,
\end{split}
\label{NE game 1}
\end{equation}
In the non-cooperative game, the optimization problem (\ref{NE game 1}) is repeatedly done by both nodes until an equilibrium is reached, if any.

If deriving the Pareto boundary straight from problem (\ref{non-convex boundary problems}), one needs to simultaneously search $\boldsymbol{Q}_1$ and $\boldsymbol{Q}_2$ in $\mathcal{X}_1\times\mathcal{X}_2$. In contrast, the non-cooperative game for an NE is in a fully distributed fashion, where each node derives its own $\boldsymbol{Q}_i$ from $\mathcal{X}_i$. More important, problem (\ref{non-convex boundary problems}) is non-convex while problem (\ref{NE game 1}) is convex. Accordingly, an NE promises much higher computational efficiency than the Pareto boundary. An NE is of the competitive optimality, however, not guaranteed to achieve the Pareto optimality. Therefore, the tradeoff between performance and computational efficiency should be considered in the strategy design for a FD channel.

In the sequel, we will investigate further into the transmission strategy design within the framework of game theory so as to improve the performance of the two-way FD channel. 
We will first consider the simple case where each FD node is equipped with single receive antenna, and then consider the general MIMO case.

\section{MISO Full-duplex Channel}
We consider the scenario where all FD nodes are equipped with only one receive antenna i.e., $N=1$. Consequently, all the channels are reduced to MISO, and can be represented by vectors $\boldsymbol{h}_{ij}, i,j\in\{1,2\}$. The maximum rate for the channel from node $i$ to node $j$ can be then simplified as
\begin{equation}
R_i(\boldsymbol{Q}_i,\boldsymbol{Q}_j) =  
 \log \Bigg(
1+\frac{\eta_{ij}\boldsymbol{h}_{ij}^{\dag}\boldsymbol{Q}_i\boldsymbol{h}_{ij}}{1+\beta\eta_{jj}\boldsymbol{h}_{jj}^{\dag}\text{diag}
(\boldsymbol{Q}_j)\boldsymbol{h}_{jj}}\Bigg),
\label{MISO achieveble rate}
\end{equation}
where $(\boldsymbol{Q}_{1},\boldsymbol{Q}_2)$ are the given transmit covariance matrices. 
In contrast with the general MIMO rates in (\ref{R1}) and (\ref{R2}), the rate for the MISO case in (\ref{MISO achieveble rate}) is in a simpler form, which then improves the efficiency in solving the Pareto boundary.

\subsection{Decoupled Optimization Problems}

The difficulty in deriving Pareto boundary for the MIMO FD channel is caused by the non-convexity and the coupled high-dimensional nature of problem (\ref{non-convex boundary problems}). To render the derivation tractable, we need to decouple problem (\ref{non-convex boundary problems}) in terms of lower-dimensional variables. Inspired by the decoupling procedure in \cite{Shang2007Asilomar,Shang2011ITInf,zhang2010ITSP} we introduce an auxiliary variable $z_i$ to denote the power of the received signal at node $j$ i.e., $z_i\triangleq{\boldsymbol{h}_{ij}^{\dag}\boldsymbol{Q}_i\boldsymbol{h}_{ij}}$. With $z_i$, we then construct the following optimization problem of $\boldsymbol{Q}_i$ under the transmit power constraint $P_i$: 
\begin{equation}
\begin{split}
\min\;\;& 
{\boldsymbol{h}_{ii}^{\dag}\text{diag}(\boldsymbol{Q}_i)\boldsymbol{h}_{ii}}\\
\text{subject to}\;\;& {\boldsymbol{h}_{ij}^{\dag}\boldsymbol{Q}_i\boldsymbol{h}_{ij}}=z_i\\
& \textbf{tr}(\boldsymbol{Q}_i)\leq P_i, \boldsymbol{Q}_i\succeq 0
\end{split}
\label{region split problem1}
\end{equation}
where $i,j\in\{1,2\}$ and $i\neq j$. Here, we require 
\begin{equation}
 0\leq z_i\leq\max\limits_{\boldsymbol{Q}_i\in\mathcal{X}_i}\boldsymbol{h}_{ij}^{\dag}\boldsymbol{Q}_i\boldsymbol{h}_{ij}=P_i\|\boldsymbol{h}_{ij}\|^2
 \label{range of z}
\end{equation}
so that problem (\ref{region split problem1}) always has a feasible solution. We define $\overline{\mathcal{R}}$ as the set of all optimal solutions for problem (\ref{region split problem1}). 

Unlike problem (\ref{non-convex boundary problems}) where the objective function is regrading $\boldsymbol{Q}_1$ and $\boldsymbol{Q}_2$, problem (\ref{region split problem1}) depends only on $\boldsymbol{Q}_i$. In Lemma \ref{Lemma R and tildeR}, we show that the Pareto boundary for the MISO FD channel can be alternatively characterized by solving problem (\ref{region split problem1}). 

\begin{lem} 
For a MISO full-duplex channel with the transmit power constraint $P_i$, any point on the Pareto boundary $\mathcal{R}^*$ for the achievable rate region $\mathcal{R}$ in (\ref{centralized region}) can be achieved by the optimal solution $\boldsymbol{Q}_i^*$ for problem (\ref{region split problem1}) with some $z_i$. That is to say, $\mathcal{R}^*\subseteq\overline{\mathcal{R}}$.
\label{Lemma R and tildeR} 
\end{lem}
\begin{proof}
Denote the optimal value of problem (\ref{region split problem1}) as $\Gamma_i^*(z_i)$. Then, we can define $\overline{R}$ in terms of $z_i$ and $\Gamma_i^*(z_i)$ as follows:
\begin{equation}
\overline{\mathcal{R}}\triangleq\bigcup_{
\scriptstyle z_1\in[0,P_1\|\boldsymbol{h}_{12}\|^2],
\atop
\scriptstyle z_2\in[0,P_2\|\boldsymbol{h}_{21}\|^2]}
\left\lbrace
\begin{split}
&
(r_1,r_2):\\
& r_1= \log \Bigg(
1+\frac{\eta_{12}z_2}{1+\beta\eta_{11}\Gamma_1^*(z_1)}\Bigg)\\
& r_2=\log \Bigg(
1+\frac{\eta_{21}z_1}{1+\beta\eta_{22}\Gamma^*_2(z_2)}\Bigg)
\end{split}
\right\rbrace. 
\nonumber   
 \label{convex boundary} 
\end{equation}

For any point $(R_1^*,R_2^*)$ on the Pareto boundary, assume that it is achieved by $\boldsymbol{Q}_1^*$ and $\boldsymbol{Q}_2^*$. $\boldsymbol{Q}_i^*$ is a feasible solution for problem (\ref{region split problem1}) with $z_i=z_i^*=\boldsymbol{h}_{ij}^{\dag}\boldsymbol{Q}_i^*\boldsymbol{h}_{ij}$ where $i,j\in\{1,2\}$ and $i\neq j$. Let $i=1$, if $\boldsymbol{Q}_1^*$ is not an optimal solution for problem (\ref{region split problem1}) i.e., $\boldsymbol{h}_{11}^{\dag}\text{diag}(\boldsymbol{Q}_1^*)\boldsymbol{h}_{11}>\Gamma_1^*(z_1^*)$ then 
\begin{equation}
R_1^*<\log \Bigg(
1+\frac{\eta_{21}z_2^*}{1+\beta\eta_{11}\Gamma_1^*(z_1^*)}\Bigg)=\overline{R_1},
\nonumber
\end{equation}
while
\begin{equation}
R_2^*\leq\log \Bigg(
1+\frac{\eta_{12}z_1^*}{1+\beta\eta_{22}\Gamma_2^*(z_2^*)}\Bigg)=\overline{R_2}.
\nonumber
\end{equation}
As $(\overline{R_1},\overline{R_2})$ belongs to $\overline{\mathcal{R}}$ and thus belongs to $\mathcal{R}$, $R_1^*<\overline{R_1}$ and $R_2^*\leq\overline{R_2}$ contradict to the Pareto optimality of $(R_1^*,R_2^*)$. Therefore $\boldsymbol{Q}_1^*$ is an optimal solution for problem (\ref{region split problem1}). In the same way we can show that $\boldsymbol{Q}_2^*$ is an optimal solution for problem (\ref{region split problem1}).
\end{proof}

We stress that the set $\overline{\mathcal{R}}$ is not necessarily equivalent to the Pareto boundary $\mathcal{R}^*$, since $\overline{\mathcal{R}}$ may include the rate pairs inside the region $\mathcal{R}$. However,  the relationship $\mathcal{R}^*\subseteq\overline{\mathcal{R}}$ implies that any approach of obtaining the set $\overline{\mathcal{R}}$ will suffice to derive the entire Pareto boundary $\mathcal{R}^*$. Furthermore, any result applying to $\overline{\mathcal{R}}$ also works for $\mathcal{R}^*$. Hence, we proceed to explore the optimal signaling for the MISO FD two-way channel by the study of the set $\overline{\mathcal{R}}$.  


\subsection{Optimal Beamforming}
Problem (\ref{region split problem1}) is not a common optimization problem since the objective function includes the non-linear operator $\text{diag}(\cdot)$. By setting $\boldsymbol{A}_i=\boldsymbol{h}_{ij}\boldsymbol{h}_{ij}^\dag$, $\boldsymbol{C}_i=\text{Diag}(|\boldsymbol{h}_{ii}^{(1)}|^{2},\dots,|\boldsymbol{h}_{ii}^{(M)}|^{2})$ and using the equivalent relationship
$\boldsymbol{h}_{ii}^{\dag}\text{diag}(\boldsymbol{Q}_i)\boldsymbol{h}_{ii}=\textbf{tr}(\boldsymbol{C}_i\boldsymbol{Q}_i)$, we reformulate problem (\ref{region split problem1}) to the semi-definite programming (SDP) problem as follows (See more details about SDP in \cite{boyd2009convex}):
\begin{equation}
\begin{split}
\min\limits_{\boldsymbol{Q}_i}\;\;& {\textbf{tr}(\boldsymbol{C}_i\boldsymbol{Q}_i)}\\
\text{subject to}\;\;&\textbf{tr}(\boldsymbol{A}_i\boldsymbol{Q}_i)=z_i,\boldsymbol{Q}_i\in\mathcal{X}_i,
\end{split}
\label{two node SDP}
\end{equation}
where $\boldsymbol{C}_i, \boldsymbol{A}_i\in \mathbb{H}^M$.
The above SDP reformulation reveals the hidden convexity of problem (\ref{region split problem1}) so that we can solve it by employing the well-developed interior-point algorithm within polynomial time. Furthermore, we can numerically characterize the Pareto boundary for the MISO FD two-way channel in efficiency.

The optimal solutions for problem (\ref{two node SDP}) determine the signaling structure to achieve the rate pairs in the set $\overline{\mathcal{R}}$. In Theorem \ref{Theorem optimality of beamforming}, we explore the rank of optimal solutions $\boldsymbol{Q}^*_i$ for problem (\ref{two node SDP}) where $i,j\in\{1,2\}$ and $i\neq j$.

\begin{thm}
For problem (\ref{two node SDP}) with $P_i \geq 0$ and  $0\leq z_i\leq P_i\|\boldsymbol{h}_{ij}\|^2$, there always exists an optimal solution $\boldsymbol{Q}_i^*$  with $\text{rank}(\boldsymbol{Q}^*_i)=1$.
\label{Theorem optimality of beamforming}
\end{thm}

\begin{proof}
See Appendix A.
\end{proof}

Note that the transmit signal with the rank-one covariance matrix can be implemented by transmitter beamforming. It follows from Theorem \ref{Theorem optimality of beamforming} that all points in the set $\overline{\mathcal{R}}$, which include the entire Pareto boundary, can be achieved by the transmitter beamforming. Therefore, we conclude that transmitter beamforming is an optimal scheme for the MISO FD two-way channel. In Lemma \ref{Lemma optimal beamforming weights} we derive the closed-form of the optimal weights for transmitter beamforming.

\begin{lem}
For node $i$ in the MISO point-to point FD wireless network with the transmit power constraint $P_i$ and complex channels $\boldsymbol{h}_{ii}, \boldsymbol{h}_{ij}, i,j\in\{1,2\}, i\neq j$, the optimal beamforming weights have the following form:

\begin{equation}
\boldsymbol{w}_i^*=\frac{\sqrt{z_i}(\boldsymbol{C}_i+\epsilon\boldsymbol{I})^{-1}\boldsymbol{h}_{ij}}{\boldsymbol{h}_{ij}^\dag(\boldsymbol{C}_i+\epsilon\boldsymbol{I})^{-1}\boldsymbol{h}_{ij}}
\label{optimal bf}
\end{equation}
where $\boldsymbol{C}_i=\text{Diag}(|\boldsymbol{h}_{ii}^{(1)}|^{2},\dots,|\boldsymbol{h}_{ii}^{(M)}|^{2})$, constant $z_i$ is within the range $0\leq z_i\leq P_i\|\boldsymbol{h}_{ij}\|^2$ and $\boldsymbol{I}$ denotes the $M\times M$ identical matrix. For a fixed $z_i$, nonnegative constant $\epsilon$ is adjusted to satisfy the transmit power constraint $\|\boldsymbol{w}_i\|^2\leq P_i$. Specially, $\epsilon=0$ if
\begin{equation}
 z_i\leq \frac{P_i(\boldsymbol{h}_{ij}^{\dag}\boldsymbol{C}_i^{-1}\boldsymbol{h}_{ij})^2}{\boldsymbol{h}_{ij}^{\dag}\boldsymbol{C}_i^{-2}\boldsymbol{h}_{ij}}.
 \label{low zi condition}
\end{equation}
\label{Lemma optimal beamforming weights} 
\end{lem}

\begin{proof}
The optimal beamforming weights can be obtained by solving problem (\ref{two node SDP}) with the rank-one constraint $\boldsymbol{Q}_i=\boldsymbol{w}_{i}\boldsymbol{w}_{i}^{\dag}$ as follows:
\begin{equation}
\begin{split}
\min\limits_{\boldsymbol{w}_i}\;\;&{\boldsymbol{w}_{i}^{\dag}\boldsymbol{C}_i\boldsymbol{w}_{i}} \\
\text{subject to}\;\;&|\boldsymbol{w}_{i}^{\dag}\boldsymbol{h}_{ij}|^2=z_i, \|\boldsymbol{w}_{i}\|^2\leq P_i.
\end{split}
\label{minimizaing BF problem1}
\end{equation}
 The above problem has the general closed-form optimal solution (\ref{optimal bf}) (see details in \cite{cox1987ITASP}). Without the transmit power constraint $\|\boldsymbol{w}_{i}\|^2\leq P_i$, problem (\ref{minimizaing BF problem1}) has the following optimal solution (shown in \cite{cox1987ITASP})
 \begin{equation}
 \boldsymbol{w}_i^*=\frac{\sqrt{z_i}\boldsymbol{C}_i^{-1}\boldsymbol{h}_{ij}}{\boldsymbol{h}_{ij}^\dag\boldsymbol{C}_i^{-1}\boldsymbol{h}_{ij}}.
 \label{epsilon=0}
 \end{equation}
 Combining (\ref{epsilon=0}) and the condition (\ref{low zi condition}), 
\begin{equation}
\|\boldsymbol{w}_{i}^*\|^2=\frac{z_i\boldsymbol{h}_{ij}^{\dag}\boldsymbol{C}_i^{-2}\boldsymbol{h}_{ij}}{(\boldsymbol{h}_{ij}^{\dag}\boldsymbol{C}_i^{-1}\boldsymbol{h}_{ij})^2}\leq P_i.
\nonumber
\end{equation} 
Hence, we conclude that $\epsilon=0$ under the condition (\ref{low zi condition}).
\end{proof}
We remark that the optimal beamforming weights for node $i$ is closely parallel to the direct channel $\boldsymbol{h}_{ij}$, beamforming the signal of interest at node $j$. While the transmit front-end noise corresponding to the stronger self-interference channel is largely suppressed via the matrix $(\boldsymbol{C}_i+\epsilon\boldsymbol{I})^{-1}$.

\section{MIMO Full-duplex Channel}  
In the MIMO full-duplex network, i.e., $M>1$ and $N>1$, there is in general no approach to decouple and convexify problem (\ref{non-convex boundary problems}). To characterize the Pareto boundary, one needs to solve problem (\ref{non-convex boundary problems}) for all possible weights $(\mu_1, \mu_2)$. For each pair $(\mu_1, \mu_2)$, the optimal solutions for problem (\ref{non-convex boundary problems}) can be obtained by an exhaustive search over $(\boldsymbol{Q}_1,\boldsymbol{Q}_2)$. But the computational complexity of the exhaustive search is prohibitively high since the search is coupled by high-dimensional $\boldsymbol{Q}_1$ and $\boldsymbol{Q}_2$. In \cite{Day2012ITSP}, the numerical methods, such as Gradient Projection, are used to improve the computational efficiency. However, any numerical method can not be guaranteed to find the global optimum due to the non-convexity of problem (\ref{non-convex boundary problems}). In addition, problem (\ref{non-convex boundary problems}) can not be decoupled, implying that an extra central node is required to solve the Pareto-optimal solutions. The central node needs to have full knowledge of the FD network, which poses an extra difficulty in the implementation of a FD channel. Consequently, for the general MIMO FD channel we restrict our attention to the non-cooperative game, by which the FD channel can be operated in its competitive optimality. Such a game is convex and in a fully distributed fashion, rendering the computation tractable. At each iteration of the game, node $i$ selfishly optimizes its own performance by changing its transmit strategy $\boldsymbol{Q}_i$. The objective is to achieve the Nash equilibrium, where each node's transmit strategy is a best response to the other node's strategy.

\subsection{Existence of Nash Equilibrium}
To obtain the Nash equilibrium (NE) for a FD channel, node $i$ needs to maximize its rate $R_i$ by solving problem (\ref{NE game 1}). The feasible set of problem (\ref{NE game 1}) is $\mathcal{X}_i=\left\lbrace\boldsymbol{Q}_i\in\mathbb{H}^{2}|\boldsymbol{Q}_i\succeq0,\textbf{tr}\left(\boldsymbol{Q}_i\right)\leq P_i\right\rbrace$. We denote the optimal solution of problem (\ref{NE game 1}) as $\mathcal{B}_i(\boldsymbol{Q}_j)$, where $\mathcal{B}_i(\cdot)$ is a function of $\boldsymbol{Q}_j$ and $\mathcal{B}_i(\cdot):\mathcal{X}_j \mapsto\mathcal{X}_i$. If $\boldsymbol{Q}_j$ is given, then $\mathcal{B}_i(\boldsymbol{Q}_j)$ satisfies
\begin{equation}
R_i(\mathcal{B}_i(\boldsymbol{Q}_j),\boldsymbol{Q}_j)\geq R_i(\boldsymbol{Q}_i, \boldsymbol{Q}_j), \forall\boldsymbol{Q}_i\in \mathcal{X}_i.
\label{Best-responce optimality}
\end{equation}
 Thus, $\mathcal{B}_i(\cdot)$ is called \textit{Best-Response} function  \cite{palomar2010convex}. We then construct a mapping $\boldsymbol{\Phi}$ from the \textit{Best-Response} function $\mathcal{B}_i(\cdot)$:
\begin{equation}
\boldsymbol{\Phi}(\boldsymbol{Q}_1,\boldsymbol{Q}_2)=\left( \mathcal{B}_1(\boldsymbol{Q}_2), \mathcal{B}_2(\boldsymbol{Q}_1)\right), 
\label{strategy mapping}
\end{equation}
where $\boldsymbol{\Phi}:\mathcal{X}_1 \times\mathcal{X}_2\mapsto\mathcal{X}_1\times\mathcal{X}_2$. The input and output of $\boldsymbol{\Phi}$ are two sets of feasible transmission strategies for the FD channel, where the output strategy for node $i$ is the best response to the input strategy for node $j$. At the fixed point of $\boldsymbol{\Phi}$, the input strategies are equal to the output strategies
\begin{equation}
(\boldsymbol{Q}_1,\boldsymbol{Q}_2)=\boldsymbol{\Phi}(\boldsymbol{Q}_1,\boldsymbol{Q}_2).
\label{Definition fixed-point}
\end{equation}
It follows from (\ref{Best-responce optimality}) and (\ref{strategy mapping}) that the competitive optimality in Definition (\ref{Definition competitive optimality}) is achieved at the fixed point. Hence, for a FD channel, a NE is equivalent to a fixed-point of the mapping $\boldsymbol{\Phi}$. It follows that the NE can be achieved for the FD channel by deriving the fixed-point of the mapping $\boldsymbol{\Phi}$ in (\ref{strategy mapping}).

Unlike the Pareto boundary, which is the outer bound of the achievable rate region and thus always exists, the existence of Nash equilibrium is not obvious. In Lemma \ref{Lemma Existence of fixed-point}, we prove that a FD channel always has a Nash equilibrium, regardless of transmit power constraints and channel realizations.

\begin{lem}[Existence of NE]
There always exists at least one Nash equilibrium for any MIMO two-way full-duplex channel. That is, the mapping $\boldsymbol{\Phi}$ in (\ref{strategy mapping}) has at least one fixed point.
\label{Lemma Existence of fixed-point}
\end{lem} 
\begin{proof}
See Appendix C.
\end{proof}

Lemma \ref{Lemma Existence of fixed-point} illustrates that any FD channel has at least one Nash equilibrium. Thus, it demonstrates that the NE can be considered as an applicable performance metric for MIMO FD two-way channels. 

\subsection{Uniqueness of Nash Equilibrium}
 
 Unlike the Pareto boundary having infinitely many points, a FD channel need not necessarily have multiple Nash equilibria. One example is the MISO FD channel. In the MISO case, the Nash equilibrium is achieved by the beamforming matrix $\boldsymbol{Q}_i^{NE}=\boldsymbol{w}_i\boldsymbol{w}_i^{\dag}$, where $\boldsymbol{w}_i=\frac{\sqrt{P_i}\boldsymbol{h}_{ij}}{\boldsymbol{h}_{ij}^\dag\boldsymbol{h}_{ij}}$. Note that $(\boldsymbol{Q}_1^{NE}, \boldsymbol{Q}_2^{NE})$ depend only on the channel matrices and transmit power constraints, implying that the Nash equilibrium is unique for the MISO FD channel. It is then natural to ask conditions to guarantee the uniqueness of Nash equilibrium in a general MIMO FD channel. We denote the rank of matrix $\boldsymbol{H}_{ij}$ as $r_{ij}$ i.e., $r_{ij}\triangleq\text{rank}(\boldsymbol{H}_{ij})$. Thus we have $r_{ij}\leq \min(M,N)$. We start by assuming the direct channel matrices $\{\boldsymbol{H}_{ij}\}_{i,j\in\{1,2\},i\neq j}$ are full row-rank matrices i.e., $r_{ij}=N$. In this scenario, the following Lemma \ref{Lemma sufficient condition for full-rank case} offers the sufficient conditions to ensure the uniqueness of NE for the FD channel.
 
 \begin{lem}
 Assume the direct channel matrices $\{{\boldsymbol{H}_{ij}}\}_{i,j\in\{1,2\}, i\neq j}$ are with full row rank. The full-duplex channel is ensured to have a unique Nash equilibrium if
 \begin{equation}
 \rho\left(\boldsymbol{H}_{11}^{\dag} 
 \boldsymbol{H}_{21}^{-{\dag}}\boldsymbol{H}_{21}^{-1}\boldsymbol{H}_{11}\right)
 \rho\left(\boldsymbol{H}_{22}^{\dag} 
 \boldsymbol{H}_{12}^{-{\dag}}\boldsymbol{H}_{12}^{-1}\boldsymbol{H}_{22}\right)<\frac{\gamma_{1}\gamma_{2}}{\beta^2},
 \label{contractive condition}
 \end{equation}
 where $\rho(\boldsymbol{X})$ denotes the spectral radius of the matrix $\boldsymbol{X}$.
 \label{Lemma sufficient condition for full-rank case}
 \end{lem}
 
 \begin{proof}
 See Appendix D.
 \end{proof}

 To give the additional physical interpretation of Lemma \ref{Lemma sufficient condition for full-rank case}, assume $\mathcal{C}(x)$ to be the cumulative distribution function (cdf) of $\rho\left(\boldsymbol{H}_{11}^{\dag} 
  \boldsymbol{H}_{21}^{-{\dag}}\boldsymbol{H}_{21}^{-1}\boldsymbol{H}_{11}\right)
  \rho\left(\boldsymbol{H}_{22}^{\dag} 
  \boldsymbol{H}_{12}^{-{\dag}}\boldsymbol{H}_{12}^{-1}\boldsymbol{H}_{22}\right)$. Following Lemma \ref{Lemma sufficient condition for full-rank case}, the Nash equilibrium is guaranteed to be unique with probability $\mathcal{C}({\eta_{21}\eta_{12}}/\beta^2\eta_{11}\eta_{22})$. Due to the non-decreasing property of cdf, $\mathcal{C}({\eta_{21}\eta_{12}}/\beta^2\eta_{11}\eta_{22})$ increases as $\eta_{21},\eta_{12}$ increases, or $\beta,\eta_{11},\eta_{22}$ decreases. Note that for a FD channel $\eta_{21},\eta_{12}$ represent the power gains of the direct channels, whereas the strength of the residual self-interference is determined by $\beta,\eta_{11},\eta_{22}$. Thus, one can increase the probability that the FD channel has a unique Nash equilibrium  by improving the direct channel gain or suppressing the residual self-interference.
 
 In Lemma \ref{Lemma sufficient condition for full-rank case}, the uniqueness of NE is guaranteed by the contractive property of the mapping $\boldsymbol{\Phi}$ with respect to the weighted-maximum norm. However, without the full-rank assumption in the above lemma, the contractive property may not hold for the mapping $\boldsymbol{\Phi}$ even if condition (\ref{contractive condition}) is satisfied. As an example, consider a symmetric FD channel with $P_1=P_2=10$, $\beta\eta_{11}/\eta_{21}=\beta\eta_{22}/\eta_{12}=1$, where the channel matrices are set as
 \begin{equation}
 \boldsymbol{H}_{11}=\boldsymbol{H}_{22}=
 \left[\begin{array}{cc}
  -0.1440+0.3203i&-0.6735-0.0040i\\-0.4009+0.5149i&-0.0351+0.6118i\\1.3155 + 0.5694i&-1.2339 - 0.4902i\\
  \end{array}\right]
 \end{equation}
 \begin{equation}
  \boldsymbol{H}_{12}=\boldsymbol{H}_{21}=
  \left[\begin{array}{cc}
   1.1187 + 0.8794i &  1.0068 - 0.0645i\\0.1281 - 0.3943i &  0.8477 + 0.3248i\\1.5970 + 0.2708i & -0.3452 + 2.3450i\\
   \end{array}\right].
  \end{equation}
 Note that the mapping $\boldsymbol{\Phi}$ is a contraction with respect to the weighted-maximum norm only if there exists some $\boldsymbol{w}=[w_1,w_2]>0$ such that 
 \begin{eqnarray}
 &&\left\|\boldsymbol{\Phi}\left(\boldsymbol{Q}_1^{(1)},\boldsymbol{Q}_2^{(1)}\right) -\boldsymbol{\Phi}\left( \boldsymbol{Q}_1^{(2)},\boldsymbol{Q}_2^{(2)}\right) \right\|_{F}^{\boldsymbol{w}}<
 \left\| 
 \left(\boldsymbol{Q}_1^{(1)},\boldsymbol{Q}_2^{(1)}\right)-\left(\boldsymbol{Q}_1^{(2)},\boldsymbol{Q}_2^{(2)}\right)
 \right\|_F^{\boldsymbol{w}}, \nonumber\\
 &&\forall\; \boldsymbol{Q}_1\in \mathcal{X}_1,\boldsymbol{Q}_2\in \mathcal{X}_2,
 \label{contractive necessary condition}
 \end{eqnarray}
 where $\mathcal{X}_i=\left\lbrace\boldsymbol{Q}\in\mathbb{H}^{2}|\boldsymbol{Q}\succeq0,\textbf{tr}\left(\boldsymbol{Q} \right)\leq 10\right\rbrace$.
 Let 
 \begin{equation}
  \boldsymbol{Q}_{1}^{(1)}=
  \boldsymbol{Q}_{1}^{(2)}=
  \left[\begin{array}{cc}
    0.2208&0\\ 0  &  9.7792\\
   \end{array}\right]
  \end{equation}
  \begin{equation}
   \boldsymbol{Q}_{2}^{(1)}=
   \boldsymbol{Q}_{2}^{(2)}=
   \left[\begin{array}{cc}
     0.4832   &   0\\  0  &  9.5168\\
    \end{array}\right].
   \end{equation}
The above set up leads to $\left\|\boldsymbol{\Phi}\left(\boldsymbol{Q}_1^{(1)},\boldsymbol{Q}_2^{(1)}\right) -\boldsymbol{\Phi}\left( \boldsymbol{Q}_1^{(2)},\boldsymbol{Q}_2^{(2)}\right) \right\|_{F}^{\boldsymbol{w}}=0.1804/\min(w_1,w_2)$ and $\left\|\left(\boldsymbol{Q}_1^{(1)},\boldsymbol{Q}_2^{(1)}\right)-\left( \boldsymbol{Q}_1^{(2)},\boldsymbol{Q}_2^{(2)}\right)
 \right\|_F^{\boldsymbol{w}}=0.1728/\min(w_1,w_2)$, implying that condition (\ref{contractive necessary condition}) is not satisfied for any $\boldsymbol{w}=[w_1,w_2]>0$. Hence, the mapping $\boldsymbol{\Phi}$ is not a contraction. However, $\rho\left(\boldsymbol{H}_{11}^{\dag} 
 \boldsymbol{H}_{21}^{-{\dag}}\boldsymbol{H}_{21}^{-1}\boldsymbol{H}_{11}\right)=0.4657<1$ and $\rho\left(\boldsymbol{H}_{22}^{\dag} 
 \boldsymbol{H}_{12}^{-{\dag}}\boldsymbol{H}_{12}^{-1}\boldsymbol{H}_{22}\right)=0.4657<1$, so condition (\ref{contractive condition}) is satisfied. The example therefore demonstrates that Lemma \ref{Lemma sufficient condition for full-rank case} is not true without the full row-rank constraints on $\boldsymbol{H}_{12},\boldsymbol{H}_{21}$. To extent the contractive property of the mapping $\boldsymbol{\Phi}$ to all FD channels, stronger conditions are needed. In Theorem \ref{Theorem conditions of unique NE}, we derive the sufficient condition for a general FD channel to have a unique NE.

\begin{thm}
A full-duplex channel has a unique NE if $\alpha_1\alpha_2<1$, where $\alpha_i$ is defined as
\begin{equation}
{\alpha_i}\triangleq
\left\lbrace 
\begin{split}
&
\frac{\beta}{\gamma_{i}}\rho\left(\boldsymbol{H}_{ii}^{\dag} 
\boldsymbol{H}_{ji}^{-{\dag}}\boldsymbol{H}_{ji}^{-1}\boldsymbol{H}_{ii}\right),\;\text{if}\; \text{rank}(\boldsymbol{H}_{ji})=N,\\
& \frac{\beta}{\gamma_{i}}
\left(1+\beta \eta _{ii}P_i\rho\left(\boldsymbol{H}_{ii}^{\dag}\boldsymbol{H}_{ii}
\right)\right)
\rho
\left(
\boldsymbol{H}_{ii}^{\dag}\boldsymbol{H}_{ii}
\right)
\rho
\left(\boldsymbol{H}_{ji}^{-\dag}\boldsymbol{H}_{ji}^{-1}
\right),\;\text{otherwise}.
\end{split}
\right. 
\label{general judge matrix}
\end{equation}
\label{Theorem conditions of unique NE}
\end{thm}

\begin{proof}
See Appendix E.
\end{proof}
 Using the inequality $\rho(\boldsymbol{A}^{\dag}\boldsymbol{X}\boldsymbol{A})\leq\rho(\boldsymbol{A}^{\dag}\boldsymbol{Y}\boldsymbol{A})$, where $\boldsymbol{Y}\succeq\boldsymbol{X}\succeq0$, we can obtain
\begin{eqnarray}
\rho\left(\boldsymbol{H}_{ii}^{\dag} 
\boldsymbol{H}_{ji}^{-{\dag}}\boldsymbol{H}_{ji}^{-1}\boldsymbol{H}_{ii}\right)
&\leq&
\rho
\left(
\boldsymbol{H}_{ii}^{\dag}\boldsymbol{H}_{ii}
\right)
\rho
\left(\boldsymbol{H}_{ji}^{-\dag}\boldsymbol{H}_{ji}^{-1}
\right)\\
&\leq&
\left(1+\beta \eta _{ii}P_i\rho\left(\boldsymbol{H}_{ii}^{\dag}\boldsymbol{H}_{ii}
\right)\right)
\rho
\left(
\boldsymbol{H}_{ii}^{\dag}\boldsymbol{H}_{ii}
\right)
\rho
\left(\boldsymbol{H}_{ji}^{-\dag}\boldsymbol{H}_{ji}^{-1}
\right).
\label{general condition strength}
\end{eqnarray}
The above equality demonstrates that the condition in Theorem (\ref{Theorem conditions of unique NE}) is stronger than the condition in Lemma (\ref{Lemma sufficient condition for full-rank case}).

Theorem \ref{Theorem conditions of unique NE} can be interpreted from two perspectives. On the one hand, assume that the channel matrices are given, Theorem \ref{Theorem conditions of unique NE} then imposes constraints on $\beta,\eta_{ij},\eta_{ii}$ to ensure the unique Nash equilibrium for the FD channel. In the case that the direct channel gain $\eta_{ij}$ is also fixed, $\alpha_1\alpha_2<1$ then indicates how small the residual self-interference must be to guarantee the uniqueness of Nash equilibrium. On the other hand, if $\beta,\eta_{jj},\eta_{ij}$ are given, Theorem \ref{Theorem conditions of unique NE} then determines the probability that the FD channel is guaranteed to have a unique Nash equilibrium. In the following Corollary \ref{Corollary circulant matrix}, we discuss a special FD channel, where all channel matrices $\{{\boldsymbol{H}_{ij}}\}_{i,j\in\{1,2\}}$ are circulant. In such a case, Theorem \ref{Theorem conditions of unique NE} can be further simplified. 

\begin{cor}
Assume that all the matrices $\{{\boldsymbol{H}_{ij}}\}_{i,j\in\{1,2\}}$ have circulant structures. The full-duplex channel is ensured to have a unique Nash equilibrium if
\begin{equation}
\max\limits_{k=1,\dots,M}\frac{|\sigma_{11}(k)|^2}{|\sigma_{21}(k)|^2}
\cdot
\max\limits_{k=1,\dots,M}\frac{|\sigma_{22}(k)|^2}{|\sigma_{12}(k)|^2}
<
\frac{\gamma_{1}\gamma_{2}}{\beta^2},
\label{circulant contractive condition}
\end{equation}
where $\sigma_{ij}(k)$ represents the $k^{th}$ eigenvalue of $\boldsymbol{H}_{ij}$. Moreover, $\{|\sigma_{ij}(k)|\}_{k=1}^{M}$ are i.i.d. Rayleigh random variables with variance $M$.
\label{Corollary circulant matrix}
\end{cor} 

The above Corollary implies that a circulant FD channel has a unique Nash equilibrium with probability $P\left(\max\limits_{k=1,\dots,M}\frac{|\sigma_{11}(k)|^2}{|\sigma_{21}(k)|^2}
\cdot
\max\limits_{k=1,\dots,M}\frac{|\sigma_{22}(k)|^2}{|\sigma_{12}(k)|^2}
<
\frac{\gamma_{1}\gamma_{2}}{\beta^2}\right)$. Furthermore, the analytical form of the probability $P$ can be derived if the channel matrices are symmetric i.e., $\boldsymbol{H}_{12}=\boldsymbol{H}_{21}$ and $\boldsymbol{H}_{11}=\boldsymbol{H}_{22}$, $\gamma_{1}=\gamma_{2}=\gamma$. For the notational convenience, denote $\Gamma(x)$ as the cdf of the ratio $A/B$, where $A$ and $B$ are two independent Rayleigh random variables. The analytical expression of $\Gamma(x)$ can be found in \cite{shakil2011record}. Note that $\{\sigma_{ij}(k)\}_{k=1}^{M}$ are independent Rayleigh random variables. It follows that $\{|\sigma_{ii}(k)|/|\sigma_{ji}(k)|\}_{k=1}^{M}$ are independently distributed with cdf $\Gamma(x)$. Under the symmetric channel assumption, the probability $P$ thus can be written in terms of $\Gamma(x)$, as follows.
\begin{eqnarray}
&& P\left(\max\limits_{k=1,\dots,M}\frac{|\sigma_{11}(k)|^2}{|\sigma_{21}(k)|^2}
\cdot
\max\limits_{k=1,\dots,M}\frac{|\sigma_{22}(k)|^2}{|\sigma_{12}(k)|^2}
<
\frac{\gamma_{1}\gamma_{2}}{\beta^2}\right)\\
&=&
P\left(\max\limits_{k=1,\dots,M}\frac{|\sigma_{11}(k)|}{|\sigma_{21}(k)|}
<
\sqrt{\frac{\gamma}{\beta}}\right)\\
&=&
\prod\limits_{k=1}^{M}
P\left(\frac{|\sigma_{11}(k)|}{|\sigma_{21}(k)|}
<
\sqrt{\frac{\gamma}{\beta}}\right)
=
\Gamma^{M}\left(\sqrt{\frac{\gamma}{\beta}}\right).
\label{circulant prob}
\end{eqnarray}

\subsection{Iterative Water-filling Algorithms}
For a given FD channel, we would like to find the transmit covariance matrices $(\boldsymbol{Q}_1,\boldsymbol{Q}_2)$ that achieve the NE under transmit power constraints. Following (\ref{Definition fixed-point}), it is equivalent to obtain a fixed-point for the mapping $\boldsymbol{\Phi}$ in (\ref{strategy mapping}). 

We operate a non-cooperative game to obtain the NE for a FD channel. In the game, the transmission strategies are iteratively updated by the mapping $\boldsymbol{\Phi}$ in (\ref{strategy mapping}). That is, node $i$ changes its strategy as $\mathcal{B}_i(\boldsymbol{Q}_j)$ at each iteration of update. Note that $\mathcal{B}_i(\boldsymbol{Q}_j)$ can be easily obtained by applying the water-filling algorithm to problem (\ref{NE game 1}) \cite{palomar2010convex}. Such a non-cooperative game is equivalent to implement the iterative water-filling algorithm (IWFA) in a fully distributed fashion \cite{shum2007convergence,scutari2009mimo}. We first assume that the IWFA is synchronous. That means, all nodes adjust their transmit covariance matrices simultaneously. Then, the transmission strategies at the $k^{\text{th}}$ iteration can be written in terms of the strategies at the $(k-1)^{\text{th}}$ iteration,
\begin{equation}
(\boldsymbol{Q}_1^{(k)},\boldsymbol{Q}_2^{(k)})=\boldsymbol{\Phi}(\boldsymbol{Q}_1^{(k-1)},\boldsymbol{Q}_2^{(k-1)}).
\end{equation}
Ideally, the IWFA converges to a Nash equilibrium at the $l^{\text{th}}$ iteration if the condition $(\boldsymbol{Q}_1^{(l)},\boldsymbol{Q}_2^{(l)})=\boldsymbol{\Phi}(\boldsymbol{Q}_1^{(l)},\boldsymbol{Q}_2^{(l)})$ is satisfied. In practice, however, we set the tolerance as a small positive number $\delta$. The stopping criterion of the IWFA is then described as
\begin{equation}
\|\boldsymbol{\Phi}(\boldsymbol{Q}_1^{(l)},\boldsymbol{Q}_2^{(l)})-(\boldsymbol{Q}_1^{(l)},\boldsymbol{Q}_2^{(l)})\|_F<\delta.
\end{equation}

To deploy the synchronous IWFA, the synchronization for all nodes is required, which poses an extra issue in the implementation of a FD channel. In an enabled FD channel, the synchronization may not be available. The nodes may delay some updates and even miss some updates. Thus, in order to be robust in such case, we propose an asynchronous version of the IWFA. To describe the possible missing updates, we denote the strategies at $k^{\text{th}}$ iteration as $(\boldsymbol{Q}_1^{\tau(k)},\boldsymbol{Q}_2^{\tau(k)})$, where 
\begin{equation}
\boldsymbol{Q}_i^{\tau(k)}=
\left\lbrace 
\begin{split}
&
\mathcal{B}_i(\boldsymbol{Q}_j^{\tau(k-1)}),\;\text{if update at $k^{th}$ iteration is succeeding},\\
& \boldsymbol{Q}_i^{\tau(k-1)},\;\text{if update at $k^{th}$ iteration is missing},\\
\end{split}
\right. 
\end{equation}
where $0\leq\tau(k)\leq k$. The asynchronous IWFA has the following stopping rule,
\begin{equation}
\left\|\left(\boldsymbol{Q}_1^{\tau(k)},\boldsymbol{Q}_2^{\tau(k)}\right)-\left(\boldsymbol{Q}_1^{\tau(k-1)},\boldsymbol{Q}_2^{\tau(k-1)}\right)\right\|_F<\delta.
\end{equation}

To investigate the convergence of IWFA in a FD channel, we only need to consider the asynchronous IWFA. Since the synchronous IWFA is a special case of the asynchronous IWFA, where $\tau(k)=k, \forall k$. In Lemma \ref{Lemma convergence of IWFA}, the sufficient condition for the convergence of the asynchronous IWFA is derived.

\begin{lem}
If a full-duplex channel satisfies the condition (\ref{general judge matrix}) in Theorem (\ref{Theorem conditions of unique NE}), then the asynchronous IWFA can converge to the unique NE from any initially feasible transmit strategies $(\boldsymbol{Q}_1^{(0)},\boldsymbol{Q}_2^{(0)})$.  
\label{Lemma convergence of IWFA}
\end{lem}

\begin{proof}
The condition in Theorem \ref{Theorem conditions of unique NE} guarantees the mapping $\boldsymbol{\Phi}$ in (\ref{strategy mapping}) to be a contraction with respect to the weighted-maximum norm. It then follows from Proposition \ref{Uniqueness of fixed-point} that the FD channel has a unique Nash equilibrium. Further, the contractive property w.r.t. $\|.\|_F^{\boldsymbol{w}}$ can be used to guarantee the asynchronous convergence.  
See the details of the proof in \cite{bertsekas1983distributed}.
\end{proof}

It follows from Lemma \ref{Lemma convergence of IWFA} that the global convergence of the asynchronous IWFA is regardless of the initial point. It implies that the unique NE solved by the asynchronous IWFA is globally asymptotically stable \cite{shum2007convergence} if the FD channel satisfies the condition in Theorem \ref{Theorem conditions of unique NE}.

\section{Numerical Results}

\subsection{Performance of MISO Full-Duplex Channel}
\begin{figure}[!t] 
\centering
\vspace{-0.1cm}
\subfloat[The achievable rate regions for the symmetric MISO FD two-way channels with $\beta=-40\;\text{dB}$, $\|\boldsymbol{h}_{12}\|=\|\boldsymbol{h}_{21}\|=1$, $P_1=P_2=1$. As plotted for comparison is the half-duplex TDMA achievable rate region.]
{\includegraphics[width=3in]{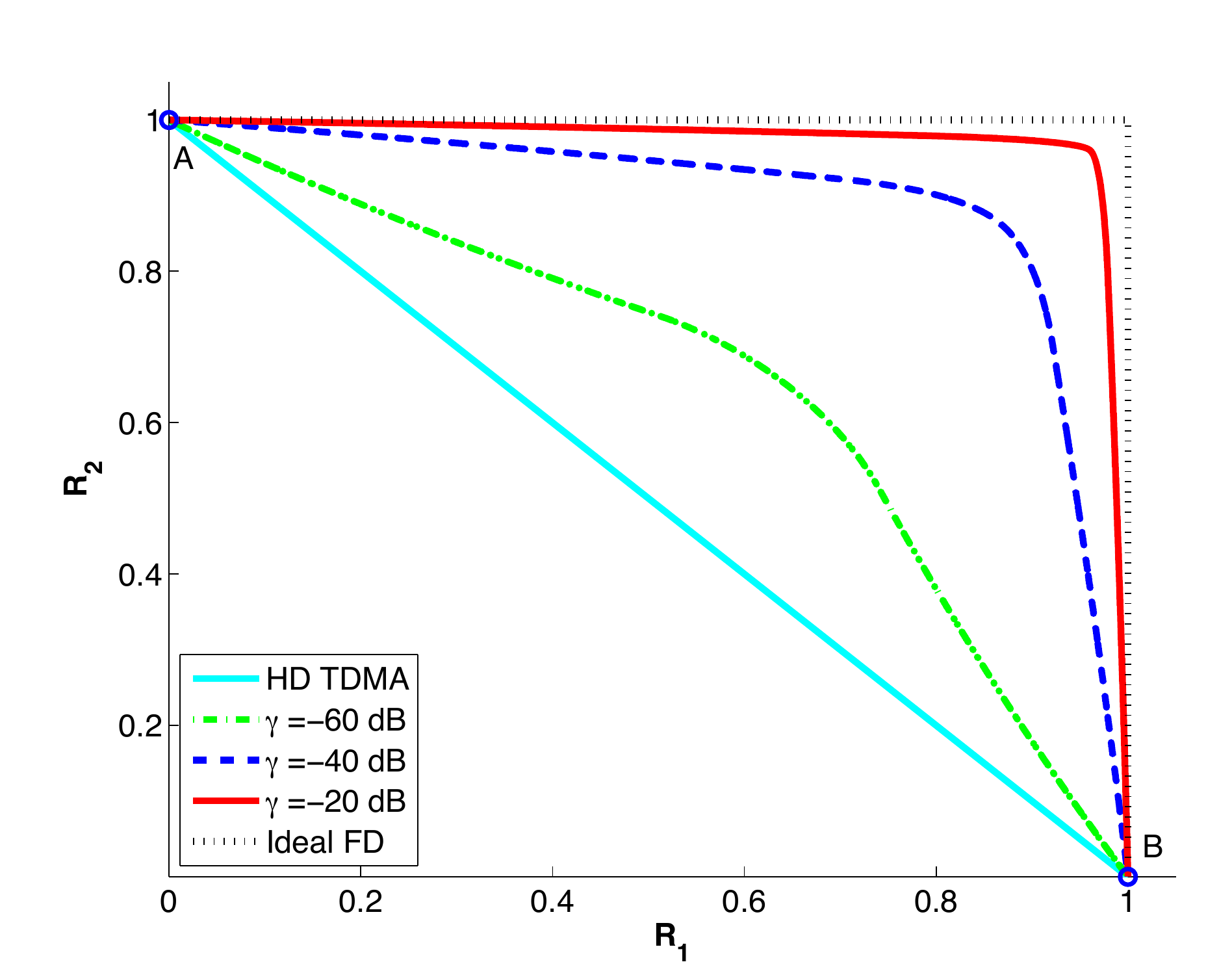}
\label{Fig4}
}
\hspace*{10pt}
\subfloat[The achievable rate regions for the symmetric MISO two-way FD channel with $\beta=-60\;\text{dB}$, $\|\boldsymbol{h}_{12}\|=\|\boldsymbol{h}_{21}\|=1$, $P_1=P_2=1$. Circles denote the ZF beamforming rates. Stars denote the NE.]
{\includegraphics[width=3in]{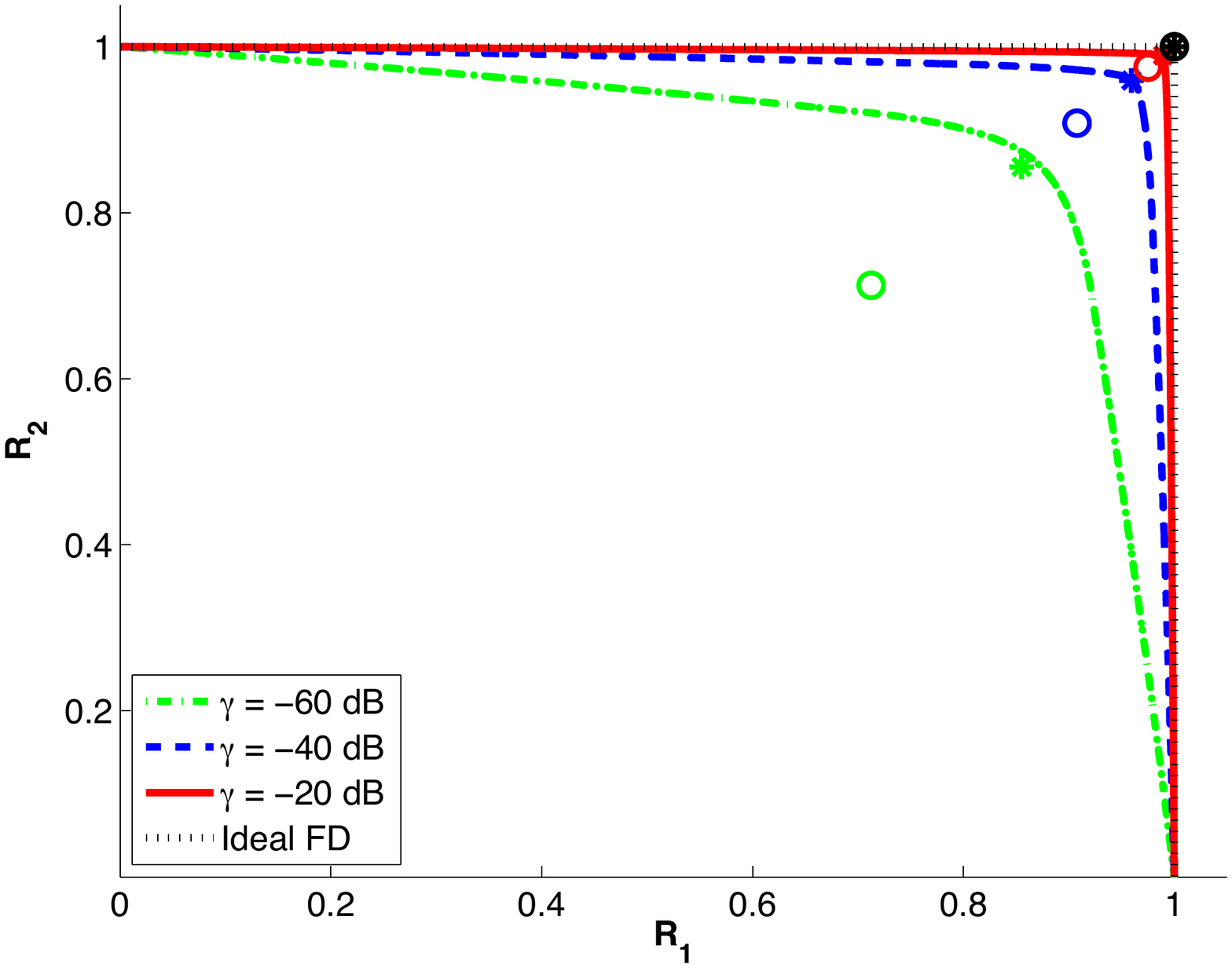}
\label{Fig6}
}
\caption{The comparison of the FD optimal beamforming, the Nash equilibrium and the zero-forcing beamforming}
\label{Figure Pareto boundary}
\vspace{-10pt}
\end{figure}

We present the achievable rate regions for the MISO full-duplex two-way channels in Fig. \ref{Fig4}, where the channels are symmetric i.e., $\boldsymbol{h}_{12}=\boldsymbol{h}_{21}$ and $\boldsymbol{h}_{11}=\boldsymbol{h}_{22}$, $\eta_{11}=\eta_{22}$ and $\eta_{12}=\eta_{21}$. Each node is equipped with $M=3$ transmit antennas and single receive antenna with the transmit power constraints $P_1=P_2=1$. And we have $\gamma_i=\eta_{ji}-\eta_{ii}$ (in dB). Here we have $\gamma_1=\gamma_2$ due to the assumption of symmetry. For the notational convenience, we use $\gamma$ to replace $\gamma_i$ in the sequel. Note that the self-interference channel gain $\eta_{ii}$ can be reduced by the passive suppression \cite{Everett2011Asilomar}, which leads to an increase of $\gamma$. The transmit front-end noise level is fixed with $\beta=-40$dB, where $\beta$ is determined by the impairments of the transmit front-end chain. With the analog and digital techniques in \cite{Duarte2012ITWC,sahai2013ITVT}, such impairments can be partly compensated so that $\beta$ can be reduced. In Fig. \ref{Figure Pareto boundary}, each colored line represents the Pareto boundary of the achievable rate region for the channel with corresponding $\gamma$. We conclude from the numerical results that the achievable rate region shrinks as $\gamma$ varies from $-20$dB to $-60$dB. However, the FD channel always outperforms than the half-duplex TDMA channel if the optimal beamforming is employed. The extreme points $A,B$ of the rate regions on the axes represent the maximum rates in the case that only one-way of the two-way channel is working. It follows that the points $A,B$ are only determined by the transmit power constraints $P_i$. The ideal MISO FD two-way channel sets the outer bound for the achievable rate regions of all channels, doubling that of the half-duplex TDMA channel.

In Fig. \ref{Fig6}, we evaluate the Pareto-boundary of the same FD channels as in Fig. \ref{Fig4} but with $\beta=-60$ dB. Comparing the channel with the same $\gamma$ in Fig. \ref{Fig4} and Fig. \ref{Fig6}, we illustrate that the achievable rate region is increased due to reduction of $\beta$. The stars in Fig. \ref{Fig6} represent the Nash equilibria for the FD channels. It can be observed that the Nash equilibrium is not Pareto-optimal except for an ideal FD channel. The circles denote the rate pairs achieved by ZF beamforming. Note that the circles are below the corresponding Nash equilibria except for the ideal FD channel. Thus, we conclude that the Nash equilibrium outperforms than the ZF beamforming for a MISO FD channel in presence of the residual self-interference.

%

\begin{figure}[!t] 
\centering
\vspace{-0.1cm}
\subfloat[The geometric comparison of the full-duplex optimal beamforming, the Nash equilibrium and the zero-forcing beamforming.]
{\includegraphics[width=3in]{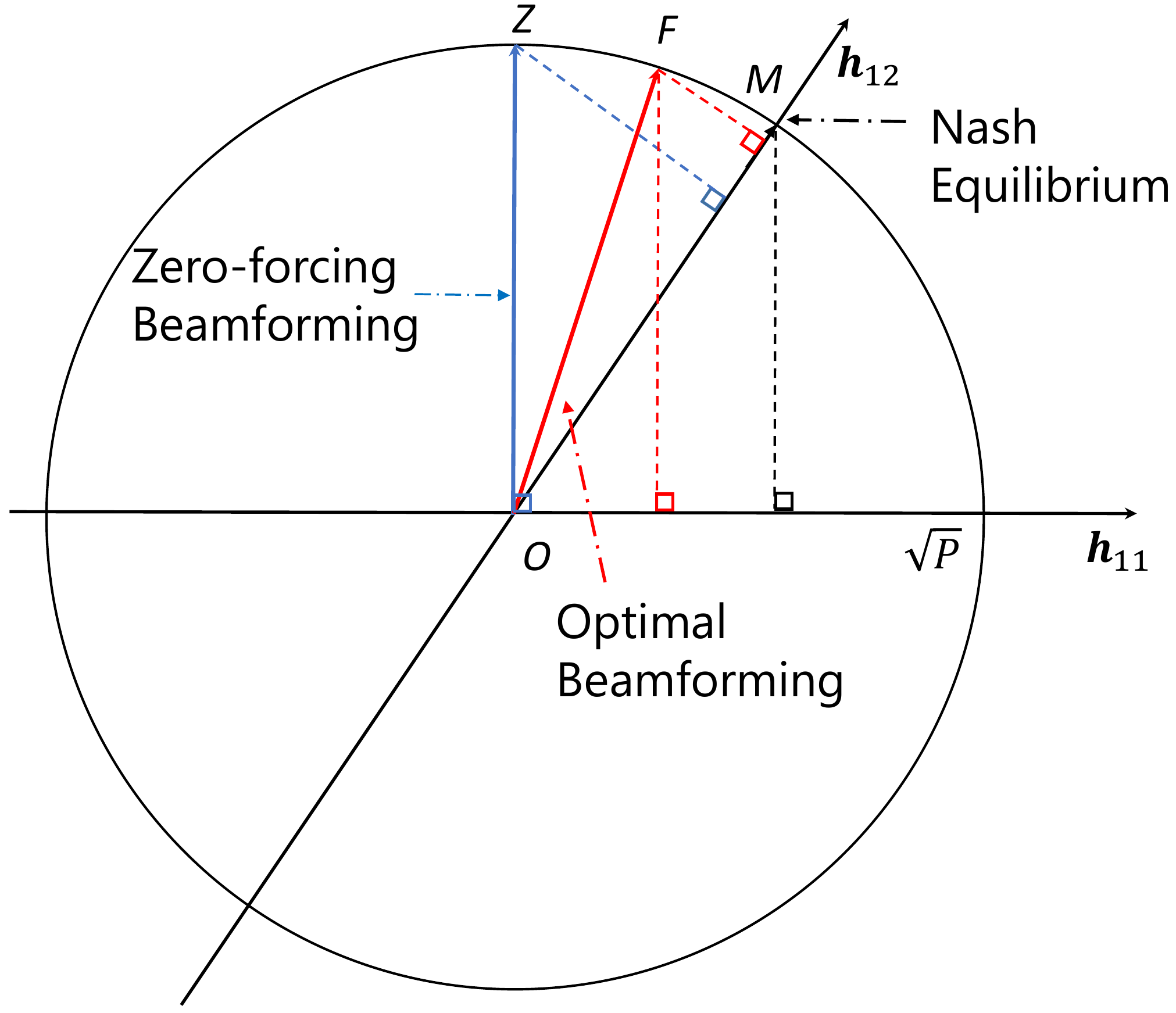}
\label{Fig8}
}
\hspace*{10pt}
\subfloat[The bit error rate of a symmetric full-duplex QPSK system. Here, $M=3$, $N=1$, $P=1$, $\beta=-60\;\text{dB}$, $\gamma=-40\;\text{dB}$.]
{\includegraphics[width=3in]{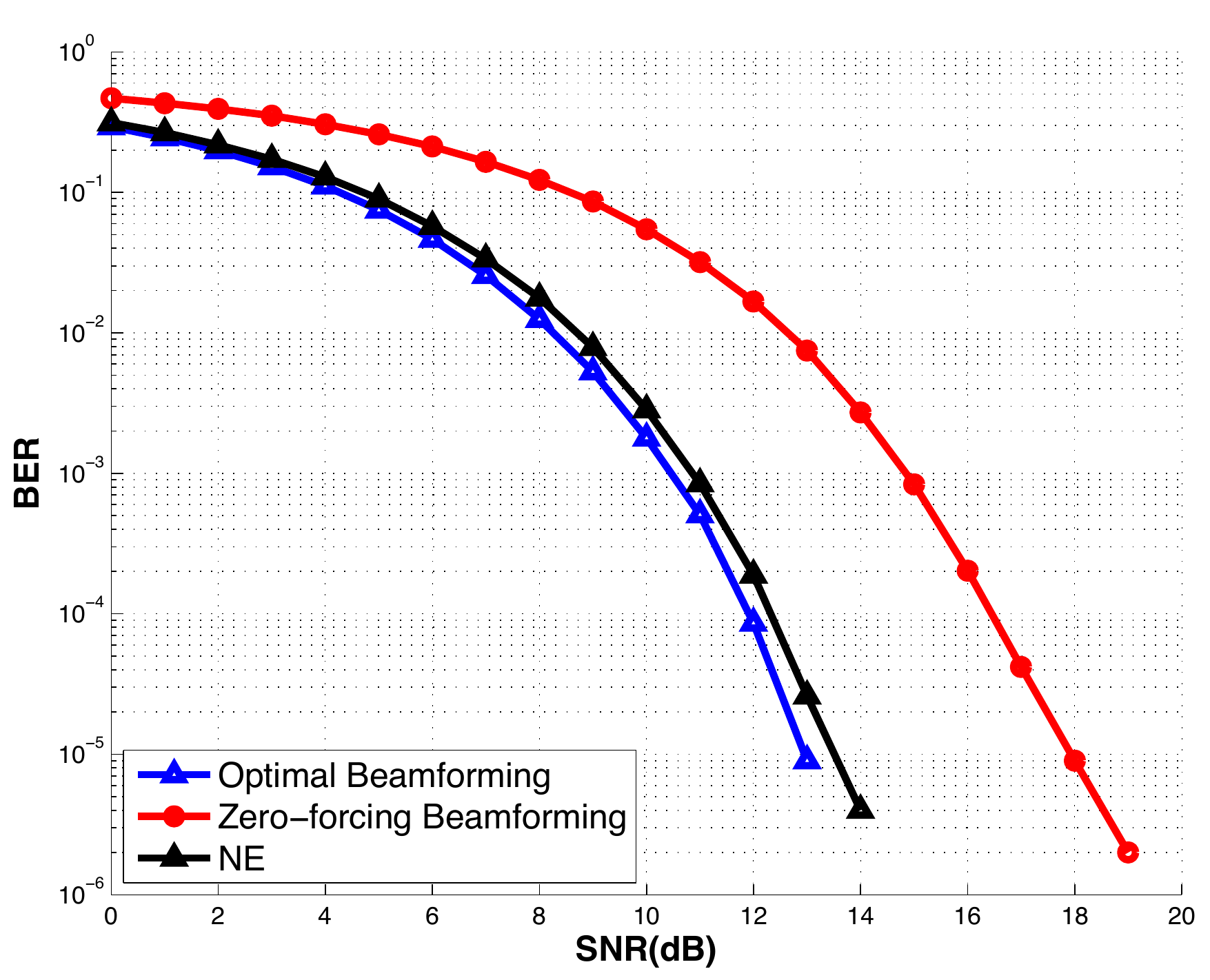}
\label{BER2}
}
\caption{}
\vspace{-20pt}
\end{figure}

In Fig. \ref{Fig8}, we geometrically compare the optimal beamforming weights, the NE and the zero-forcing beamforming weights. 
For simplicity, we assume all channels to be $2\times 1$ real-value vectors. We further assume the channels to be symmetric i.e., $\boldsymbol{h}_{11}=\boldsymbol{h}_{22}$ and $\boldsymbol{h}_{12}=\boldsymbol{h}_{21}$. Assume the transmit power constraints $P_1=P_2=P$. Then all possible beamforming weights are contained in the disc with radius $\sqrt{P}$. In Fig. \ref{Fig8}, $\overline{OF}$ represents the optimal beamforming weights by which the rate pair with maximum sum-rate can be achieved. The zero-forcing (ZF) beamforming weights is represented by $\overline{OZ}$. Note that $\overline{OZ}$ restricts the transmit signal orthogonal to the self-interference channel $\boldsymbol{h}_{ii}$. Compared with the ZF beamforming weights, the optimal $\overline{OF}$ is not orthogonal to $\boldsymbol{h}_{ii}$ but has greater length of projection on the direct channel $\boldsymbol{h}_{ij}$. Among all the weights, the Nash equilibrium $\overline{OM}$ has the greatest length of projection on $\boldsymbol{h}_{ij}$. However, the Nash equilibrium is outperformed by the optimal beamforming weights due to the larger amount of self-interference generated by $\overline{OM}$ than $\overline{OF}$. We remark that the direction of NE coincides with that of the direct channel $\boldsymbol{h}_{ij}$ in the MISO case. However, it is not true for a general MIMO FD channel.

%

The achievable rate region shown in Fig. \ref{Fig6} can be achieved only if the infinite-length Gaussian codebook is employed by all nodes. However, the communication systems in practice always choose the discrete-alphabet modulation schemes, such as QPSK or QAM. In Fig. \ref{BER2}, we show the error rate performance for one of the MISO FD channels in Fig. \ref{Fig6}, where $\gamma$ is equal to $-40$dB. We choose QPSK as the modulation scheme and employ the optimal beamforming, Nash equilibrium and zero-forcing beamforming for the FD channel respectively. Note that the FD channel is chosen to be symmetric, implying that the transmission from node $1$ to node $2$ has the same bit error rate as the transmission from node $2$ to node $1$. Therefore, we only show the BER performance for one directional transmission.


\subsection{Performance of MIMO Full-Duplex Nash Equilibrium}
\begin{figure}[!t] 
\centering
\vspace{-0.1cm}
\subfloat[Probability (lower bound) that a FD channel has the unique NE for $\beta=-40$dB, $\beta=-60$dB. Here, $M=N=3$. The channel has a circulant symmetric structure.]
{\includegraphics[width=3in]{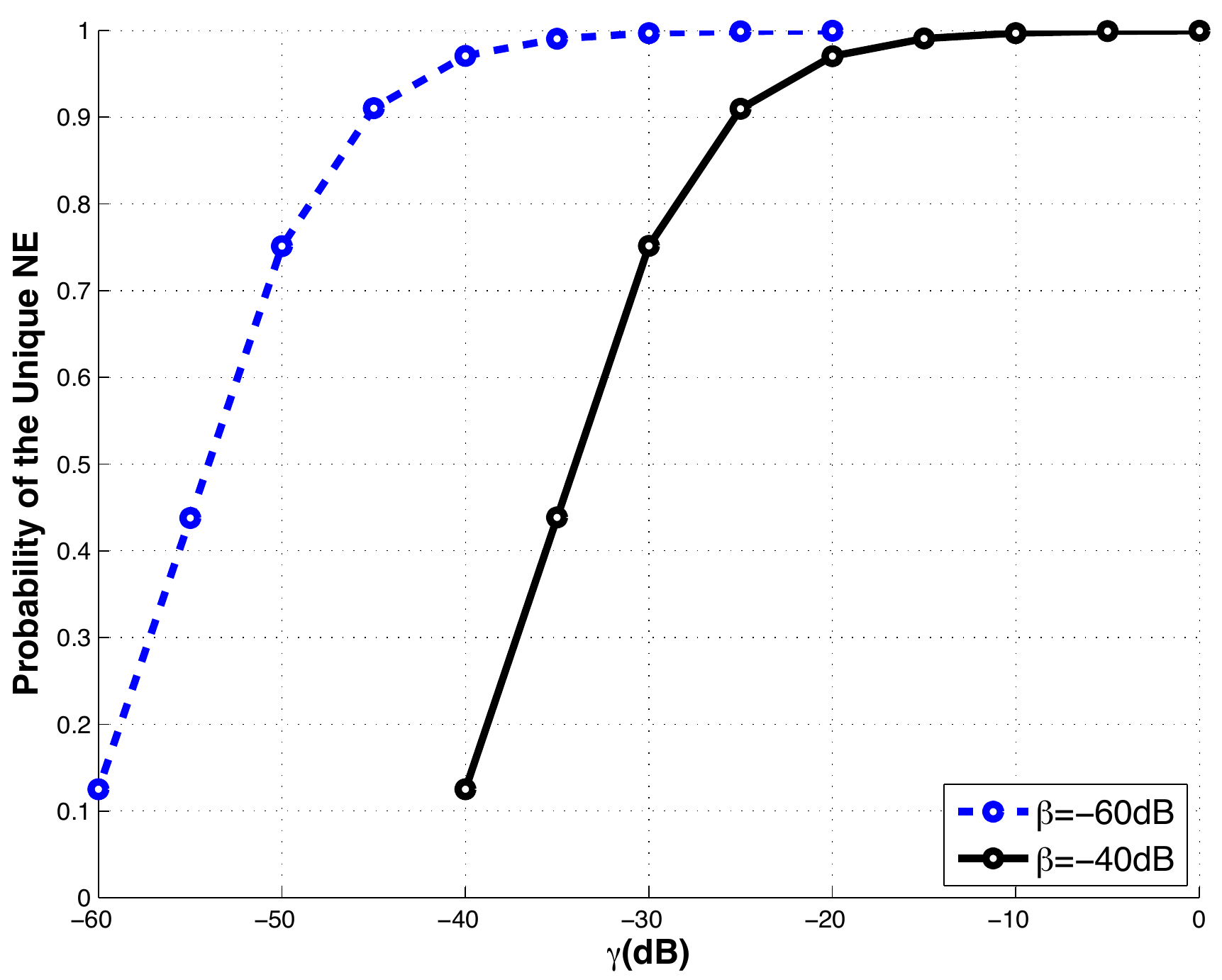}
\label{uniqueness of NE fig}
}
\hspace*{10pt}
\subfloat[Probability that the iterative water-filling algorithm converges to a Nash equilibrium in X steps. Here, $M=N=3$, $\beta=-60$dB.]
{\includegraphics[width=3in]{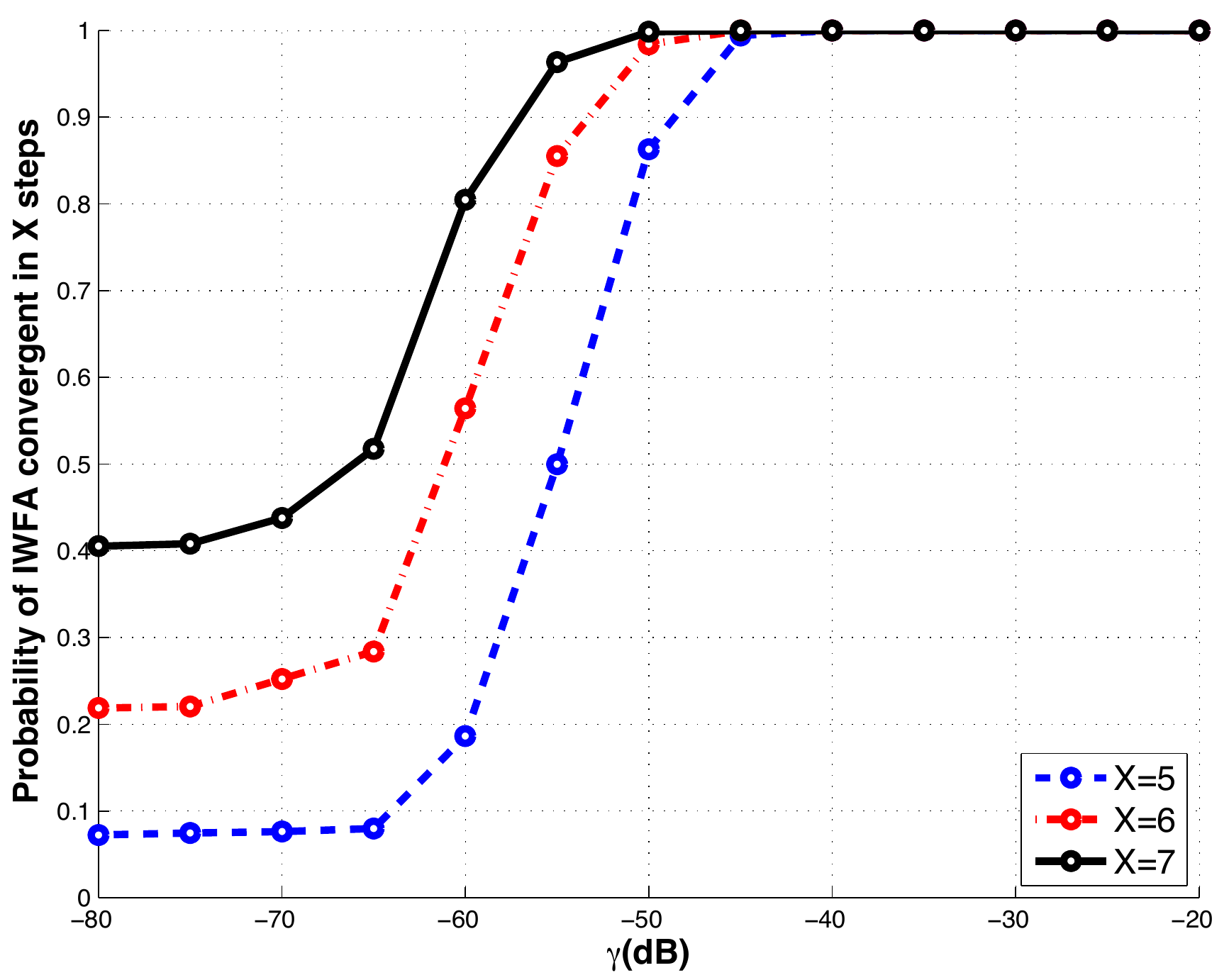}
\label{cvginX}
}
\caption{The performance of the synchronous IWFA}
\vspace*{-10pt}
\end{figure}

It follows from Theorem \ref{Theorem conditions of unique NE} that the uniqueness of NE depends on the channel realizations $\{\boldsymbol{H}_{ij}\}_{i,j\in\{1,2\}}$. With given $\beta$, $\eta_{ij}$ and $\eta_{ii}$, Theorem \ref{Theorem conditions of unique NE} then gives the probability that NE is guaranteed to be unique. Specially, for a symmetric circulant FD channel, the probability derived from Theorem \ref{Theorem conditions of unique NE} can be analytically expressed in (\ref{circulant prob}). In Fig. \ref{uniqueness of NE fig}, we plot the probability in (\ref{circulant prob}) as a function of $\gamma=\gamma_1=\gamma_2$. It follows from Lemma \ref{Lemma convergence of IWFA} that the condition in Theorem \ref{Theorem conditions of unique NE} suffices to guarantee the convergence of asynchronous IWFA. Accordingly, Fig. \ref{uniqueness of NE fig} is an analytical lower bound for the probability that IWFA can asynchronously converge to a NE. Moreover, the unique NE is globally asymptotically stable if the condition in Theorem \ref{Theorem conditions of unique NE} is satisfied. Thus, the larger probability implies that the FD NE is more stable. Accordingly, Fig. \ref{uniqueness of NE fig} indicates the stability of operating a FD network at its NE.

%

As illustrated in Fig.\ref{uniqueness of NE fig}, the probability in (\ref{circulant prob}) can be increased by either reducing $\beta$ or increasing $\gamma$. For simplicity, assume the direct channel gain is fixed. Then both the reduction of $\beta$ and the increase of $\gamma$ lead to the mitigation in residual self-interference. It implies that, for a FD channel, the stability of its competitive optimal strategies can be improved by mitigating the residual self-interference.

In Fig. \ref{cvginX}, we evaluate the computational efficiency of the synchronous iterative water-filling algorithm as a function of $\gamma$. We randomly produce 100,000 channel realizations of a $3\times3$ FD channel, testing the probability that the synchronous IWFA converges in X steps, which are then plotted in Fig. \ref{cvginX}. 

The probability that the asynchronous IWFA converges to a Nash equilibrium in X steps increases as $\gamma$ increases. It implies that the average number of iterations, which is required by IWFA to reach a Nash equilibrium, reduces with an increase of $\gamma$. Therefore, the computational efficiency of IWFA is improved if the gain of the direct channel over the self-interference channel increases.

\begin{figure}[!t] 
\centering
\includegraphics[width=3in]{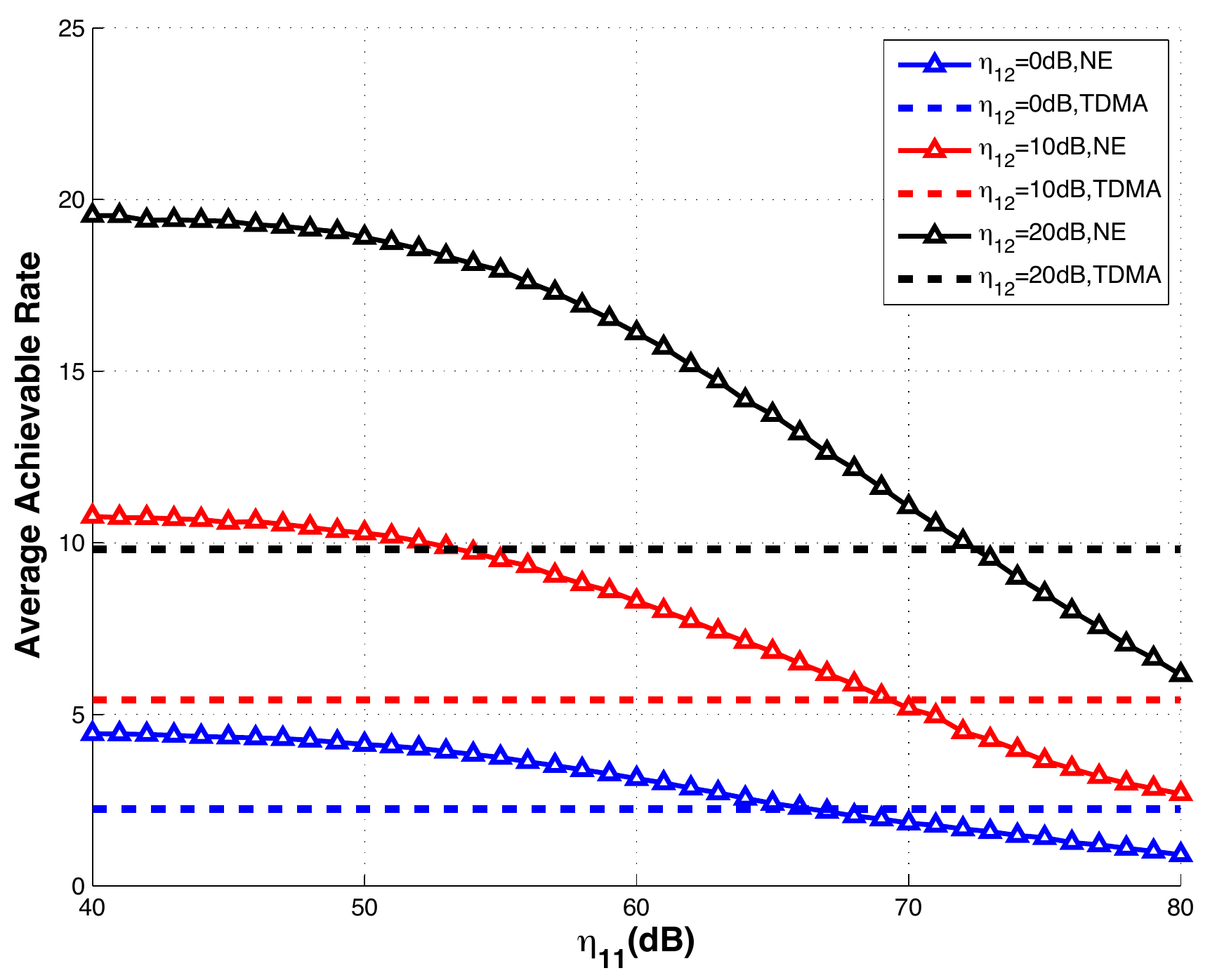}
\caption{Average achievable sum-rate for a symmetric MIMO two-way FD channel with $\beta=-60\;\text{dB}$, $P_1=P_2=10$ and $M=N=3$.} 
\label{FDVSHD}
\vspace*{-10pt}
\end{figure}

In Fig. \ref{FDVSHD}, we compare the performance of FD Nash equilibrium with half-duplex TDMA in terms of the achievable sum-rate. Each node is equipped with $M=3$ transmit antennas and $N=3$ receive antennas with the transmit power constraints $P_1=P_2=1$. The transmit front-end noise level is fixed with $\beta=-60$dB. For simplicity, the FD channel is assumed to be symmetric i.e., $\boldsymbol{H}_{12}=\boldsymbol{H}_{21}$ and $\boldsymbol{H}_{11}=\boldsymbol{H}_{22}$ with $\eta_{11}=\eta_{22}$ and $\eta_{12}=\eta_{21}$, where $\boldsymbol{H}_{ij}\sim\mathcal{CN}(\textbf{0},\boldsymbol{I}_{MN})$. The curves are averaged over 1000 independent channel realizations.  

The performances of both FD NE and half-duplex TDMA are improved as the direct channel gain $\eta_{ij}$ increases. If the self-interference channel gain $\eta_{ii}$ is less than $50$dB, then the direct channel gain $\eta_{ij}$ dominants the performance of the FD channel in terms of the achievable rate. The achievable sum-rate of FD NE is close to double that of half-duplex TDMA, implying that it is nearly optimal to operate at FD NE within the low $\eta_{ii}$ region. Beyond the low $\eta_{ii}$ region, the sum-rate of FD NE linearly decreases with the increase of $\eta_{ii}$. This is due to the fact that the residual self-interference leads the performance of the FD channel beyond the low $\eta_{ii}$ region. In contrast, TDMA works in half-duplex mode, thus the performance of TDMA is not affected by the self-interference. It follows that the gap between FD NE and half-duplex TDMA decreases as $\eta_{ii}$ increases with a given $\eta_{ij}$. Interestingly, NE and TDMA perform evenly at $\eta_{ii}=67$dB, $\eta_{ii}=69$dB and $\eta_{ii}=72$dB, if $\eta_{ij}=0$dB, $\eta_{ij}=10$dB and $\eta_{ij}=20$dB respectively. It implies that the tradeoff between FD NE and half-duplex TDMA is mainly determined by the self-interference channel gain $\eta_{ii}$ instead of the direct channel gain $\eta_{ij}$. Moreover, the achievable sum-rate of NE is smaller than that of TDMA with regardless of the value of $\eta_{ij}$ if $\eta_{ii}$ is beyond $72$dB. It follows that the NE of the FD two-way channel can bring extra benefits than the half-duplex TDMA two-way channel only if the self-interference channel gain can be suppressed to be lower than $72$dB.

\section{Conclusion}
We considered the MIMO point-to-point full-duplex wireless network under the imperfect transmit front-end chains. The network was modeled by a MIMO two-way full-duplex channel with transmit front-end noise, which was then studied from a game-theoretic view. We characterized the Pareto boundary for a MISO two-way full-duplex channel in presence of the transmit front-end noise. Using the decoupling technique and SDP reformulation, we proposed a method to obtain the entire Pareto boundary by solving a family of convex SDP problems, rather than the original non-convex problems. We showed that any rate pair on the Pareto boundary can be achieved by the beamforming transmission strategy. Moreover, we provided the closed-form solution for the optimal beamforming weights of the MISO full-duplex two-way channel. 

For the general MIMO full-duplex two-way channel, we achieved the competitive optimality by establishing the existence and the uniqueness of Nash equilibrium. We then modified the classical iterative waterfilling algorithm to efficiently reach the NE for a MIMO two-way full-duplex channel. Through our numerical results, we demonstrated that the transmit front-end noise level $\beta$ and the direct-to-self-interference channel gain ratio $\gamma$ can influence the full-duplex NE from the following perspectives. First, the computational efficiency of reaching a full-duplex NE improves with the reduction of $\beta$ or the increase of $\gamma$. Second, the stability of operating on full-duplex NE improves as $\beta$ decreases or $\gamma$ increases. Finally, we found that the self-interference channel gain $\eta_{ii}=72$dB is the threshold, beyond which the full-duplex NE is outperformed by the half-duplex TDMA in achievable rates. 

\section{Appendix}
\subsection{Proof of Theorem \ref{Theorem optimality of beamforming}}
We prove Theorem \ref{Theorem optimality of beamforming} by the primal-dual method. Note that problem (\ref{two node SDP}) is feasible and bounded. It follows that its dual problem is also feasible and bounded \cite{boyd2009convex}. Assume $\boldsymbol{Q}^*$ is an optimal solution for problem (\ref{two node SDP}). From \cite{boyd2009convex}, problem (\ref{two node SDP}) has the dual problem as follows:
\begin{equation}
\begin{split}
\min_{\lambda_1,\lambda_2}\;\;&{\lambda_1z_i+\lambda_2P} \\
\hspace{20pt}\text{subject to}\;\;&\boldsymbol{Z}=\boldsymbol{C}_i-{\lambda_1\boldsymbol{A}_i}-{\lambda_2\boldsymbol{I}}\succeq 0.\\
\end{split}
\label{equality SDP Dual}
\end{equation} 
where $P=\textbf{tr}(\boldsymbol{Q}^*)$. Assume $((\lambda_1^*,\lambda_2^*),\boldsymbol{Z}^*)$ are the optimal solutions for (\ref{equality SDP Dual}). We denote the rank of 
$\boldsymbol{Q}^*$ by $r$. We assume $r>1$. Following that $\boldsymbol{Q}^*$ is positive semi-definite, $\boldsymbol{Q}^*$ can then be written as $\boldsymbol{Q}^*=\boldsymbol{V}\boldsymbol{V}^{\dag}$ via the singular-value decomposition where $\boldsymbol{V}\in \mathbb{C}^{M \times r}$.
 
Next, we consider the following two linear equations defined by $\boldsymbol{A}_i$ and $\boldsymbol{V}$:
\begin{equation}
\left\lbrace 
\begin{split}
&\textbf{tr}(\boldsymbol{V}^{\dag}\boldsymbol{A}_i\boldsymbol{V}\boldsymbol{X})=0\\
&\textbf{tr}(\boldsymbol{X})=0\\
\end{split}
\right. 
\label{linear system}
\end{equation}  
where the unknown matrix $\boldsymbol{X} \in \mathbb{H}^r$ contains $r^2$ real-valued unknowns, that is, $\frac{r(r+1)}{2}$ for the real part and $\frac{(r(r-1))}{2}$ for the imaginary part. 

The linear system (\ref{linear system}) must have a non-zero solution, denoted by $\boldsymbol{X}^*$, since it has $r^2$ unknowns where $r\geq 2$. By decomposing the Hermitian matrix ${\boldsymbol{X}^*}$, we obtain ${\boldsymbol{X}^*}=\boldsymbol{U}\boldsymbol{\Sigma}\boldsymbol{U}^{\dag}$, where $\boldsymbol{U}$ is an $r$ dimensional unitary matrix and $\boldsymbol{\Sigma}$ is the diagonal matrix,  $\boldsymbol{\Sigma}=\text{Diag}(\sigma_1,\dots,\sigma_r)$. Without loss of generality, we assume $|\sigma_1|\geq|\sigma_2|\dots\geq|\sigma_r|$. Non-zero matrix $\boldsymbol{X}^*$ has at least one non-trivial eigenvalue, thus $|\sigma_1|>0$. Next, we construct a new matrix as follows:
\begin{equation}
{\boldsymbol{Q}^*_{(1)}}=\boldsymbol{V}(\boldsymbol{I}-\frac{1}{\sigma_1}{\boldsymbol{X}^*})\boldsymbol{V}^{\dag}.
\end{equation}
Note that $\boldsymbol{I}-\frac{1}{\sigma_1}{\boldsymbol{X}^*} \succeq 0$. It follows that $\boldsymbol{Q}^*_{(1)}$ is semi-positive definite. Next, we show that $\boldsymbol{Q}^*_{(1)}$ is also an optimal solution for problem (\ref{two node SDP}). Note that $\boldsymbol{Q}^*_{(1)}$ is optimal for problem (\ref{two node SDP}) if and only if $(\boldsymbol{Q}^*_{(1)}, (\lambda_1^*,\lambda_2^*),\boldsymbol{Z}^*)$ satisfies the KKT conditions, including the primal feasibility, the dual feasibility and the complementarity \cite{Huang2010ITSP}.
As $((\lambda_1^*,\lambda_2^*),\boldsymbol{Z}^*)$ is unchanged, the dual feasibility is automatically satisfied. Therefore, we need only to prove the primal feasibility and the complementarity of $\boldsymbol{Q}^*_{(1)}$.

$\boldsymbol{Q}^*_{(1)}$ is a feasible solution for problem (\ref{two node SDP}), since the following two equations hold for $\boldsymbol{Q}^*_{(1)}$, 
\begin{eqnarray}
\textbf{tr}(\boldsymbol{A}_i\boldsymbol{Q}^*_{(1)})
&=&\textbf{tr}(\boldsymbol{A}_i\boldsymbol{V}(\boldsymbol{I}-\frac{1}{\sigma_1}{\boldsymbol{X}^*})\boldsymbol{V}^{\dag})\\
&=&\textbf{tr}(\boldsymbol{V}^{\dag}\boldsymbol{A}_i\boldsymbol{V}\boldsymbol{I})-\frac{1}{\sigma_1}\textbf{tr}(\boldsymbol{V}^{\dag}\boldsymbol{A}_i\boldsymbol{V}{\boldsymbol{X}^*})=z_i,
\nonumber\\
\textbf{tr}(\boldsymbol{Q}^*_{(1)})
&=&\textbf{tr}(\boldsymbol{V}(\boldsymbol{I}-\frac{1}{\sigma_1}{\boldsymbol{X}^*})\boldsymbol{V}^{\dag})\\
&=&\textbf{tr}(\boldsymbol{V}^{\dag}\boldsymbol{V}\boldsymbol{I})-\frac{1}{\sigma_1}\textbf{tr}(\boldsymbol{V}^{\dag}\boldsymbol{V}{\boldsymbol{X}^*})=P.\nonumber
\end{eqnarray}

To show the complementarity, note that $\textbf{tr}(\boldsymbol{Q}^*\boldsymbol{Z}^*)=\textbf{tr}(\boldsymbol{V}^{\dag}\boldsymbol{Z}^*\boldsymbol{V})=0$ and $\boldsymbol{V}^{\dag}\boldsymbol{Z}^*\boldsymbol{V}\succeq 0$ implies that $\boldsymbol{V}^{\dag}\boldsymbol{Z}^*\boldsymbol{V}=0$. It follows that
 \begin{eqnarray}
 \textbf{tr}(\boldsymbol{Q}^*_{(1)}\boldsymbol{Z}^*)&=&\textbf{tr}(\boldsymbol{V}(\boldsymbol{I}-\frac{1}{\sigma_1}{\boldsymbol{X}}^*)\boldsymbol{V}^{\dag}\boldsymbol{Z}^*)\\
 &=&\textbf{tr}((\boldsymbol{I}-\frac{1}{\sigma_1}{\boldsymbol{X}^*})\boldsymbol{V}^{\dag}\boldsymbol{Z}^*\boldsymbol{V})=0.
 \nonumber
 \end{eqnarray}
Therefore, $\boldsymbol{Q}^*_{(1)}$ is an optimal solution for problem (\ref{two node SDP}). Furthermore, the rank of $\boldsymbol{Q}^*_{(1)}$ is strictly smaller than $r$ since $\text{rank}(\boldsymbol{Q}^*_{(1)})=\text{rank}(\boldsymbol{I}-\frac{1}{\sigma_1}{\boldsymbol{X}^*})<r$. 

We can repeat this process as $\boldsymbol{Q}^*,\boldsymbol{Q}^*_{(1)},\boldsymbol{Q}^*_{(2)},\cdots$, until $\text{rank}(\boldsymbol{Q}^*_{(k)})\leq \sqrt{2}$. In other words, the rank of the optimal solution can be strictly decreasing to $\text{rank}(\boldsymbol{Q}^*_{(k)})\leq \sqrt{2}$, that is, $\text{rank}(\boldsymbol{Q}^*_{(k)})=1$. 

\subsection{Properties of The Best-Response Function $\mathcal{B}(\cdot)$}
Introducing the constant $P_i$ and the matrices $\{\boldsymbol{H}_{ij}\}_{i,j\in\{1,2\}}$, the best-response function $\mathcal{B}_i(\boldsymbol{Q}_j)$ denotes the optimal solution of the following optimization problem:
\begin{equation}
  \begin{split}
  \max\limits_{\boldsymbol{Q}_i}\;\;& \text{log}\;\text{det}(\boldsymbol{I}+\eta_{ij}\boldsymbol{H}_{ij}^{\dag}\boldsymbol{\Sigma}_{j}^{-1}\boldsymbol{H}_{ij}\boldsymbol{Q}_i)\\
  \text{subject to}\;\;
  &\boldsymbol{Q}_i\in\mathcal{X}_i,
  \end{split} 
  \label{original waterfilling problem}
 \end{equation}  
where $i,j\in\{1,2\}$ and $i \neq j$; $\boldsymbol{\Sigma}_{i}\triangleq\boldsymbol{I}+\beta\eta_{ii}\boldsymbol{H}_{ii}\text{diag}(\boldsymbol{Q}_i)\boldsymbol{H}_{ii}^{\dag}$ which is a full-rank square matrix.

According to \cite[Lemma 11.1 and Lemma 11.2]{palomar2010convex}, problem (\ref{original waterfilling problem}) can be equivalent to 
\begin{equation}
  \begin{split}
  \min\limits_{\boldsymbol{Q}_i}\;\;& \|\boldsymbol{Q}_i-\boldsymbol{X}_0(\boldsymbol{Q}_j)\|_F\\
  \text{subject to}&\;\;
  \boldsymbol{Q}_i\in \mathcal{X}_i,
  \end{split} 
  \label{shortest distance problem}
\end{equation}
where $\mathcal{X}_i$ is a non-empty closed convex set; $\boldsymbol{X}_0(\boldsymbol{Q}_j)=-\left(\eta_{ij}\boldsymbol{H}_{ij}^{\dag}\boldsymbol{\Sigma}_{j}^{-1}\boldsymbol{H}_{ij} \right)^{-1}-c_i\boldsymbol{P}_{\mathcal{N}(\boldsymbol{H}_{ij})}$ with the positive constant $c_i$ that can be chosen independent of $\boldsymbol{Q}_j$ and the matrix $\boldsymbol{P}_{\mathcal{N}(\boldsymbol{H}_{ij})}$ that depends only on the channel matrix $\boldsymbol{H}_{ji}$. 
It follows that $\mathcal{B}_i(\boldsymbol{Q}_j)$ represents the point in $\mathcal{X}_i$ which is closest to $\boldsymbol{X}_0(\boldsymbol{Q}_j)$. The best-response function $\mathcal{B}(\cdot)$ can then be alternatively interpreted as a metric projection which projects the matrix $\boldsymbol{X}_0(\boldsymbol{Q}_j)$ onto the set $\mathcal{X}_i$. 

The interpretation of the function $\mathcal{B}(\cdot)$ as the metric projection implies the non-expansive property as follows:
\begin{equation}
\|\mathcal{B}_i(\boldsymbol{Q}_j^{(1)})-\mathcal{B}_i(\boldsymbol{Q}_j^{(2)})\|_F\leq \|\boldsymbol{X}_0(\boldsymbol{Q}_j^{(1)})-\boldsymbol{X}_0(\boldsymbol{Q}_j^{(2)})\|_F,\;\;    \forall \;\boldsymbol{Q}_j^{(1)},\boldsymbol{Q}_j^{(2)}\in \mathbb{C}^{M \times M}.
\label{non-expansive property}
\end{equation}

Building on the continuity of the metric projection and using the continuity of $\boldsymbol{X}_0(\boldsymbol{Q}_i)$ on $\boldsymbol{Q}_i$, one can obtain the continuous property of the best-response function $\mathcal{B}(\boldsymbol{Q}_i)$ on $\boldsymbol{Q}_i$.
 
\subsection{Proof of Lemma \ref{Lemma Existence of fixed-point}} 
Following (\ref{Definition fixed-point}), a Nash equilibrium is equivalent to a fixed-point of the mapping $\boldsymbol{\Phi}$ in (\ref{strategy mapping}). Thus, we will use the following results in the fixed-point theory from \cite{agarwal2001fixed,palomar2010convex} to show the existence of NE.

\begin{prop}
Given a mapping $\mathcal{T}: \mathcal{X}\mapsto\mathcal{X}$. If $\mathcal{X}$ is nonempty, convex and compact subset of a finite-dimensional normed vector space and $\mathcal{T}$ is a continuous mapping, then $\mathcal{T}$ has at least one fixed-point.
\label{existence of fixed-point}
\end{prop} 

Using Proposition \ref{existence of fixed-point}, it suffices to show the existence of fixed-points for the water-filling mapping $\boldsymbol{\Phi}$ in (\ref{strategy mapping}). The water-filling mapping $\boldsymbol{\Phi}$ in (\ref{strategy mapping}) is defined on the set $\mathcal{X}_1\times\mathcal{X}_2$, where $\mathcal{X}_i=\left\lbrace\boldsymbol{Q}\in\mathbb{H}^{M}|\boldsymbol{Q}\succeq0,\textbf{tr}\left(\boldsymbol{Q} \right)\leq P_i\right\rbrace$. It follows from the results in \cite{boyd2009convex} that $\mathcal{X}_i$ is a convex and compact subset of $\mathbb{C}^{M^2}$ for any $P_i\in\mathbb{R}$. Thus, $\mathcal{X}$ is nonempty, convex and compact set with a finite dimension for any transmit power constraints $P_1$ and $P_2$. In Appendix B, We prove the continuity of the Best-response function $\mathcal{B}_i(\cdot)$ on $\mathcal{X}_i$ for any given set of channel matrices $\{\boldsymbol{H}_{ij}\}_{i,j\in\{1,2\}}$. It follows that the mapping $\boldsymbol{\Phi}$ is continuous on $\mathcal{X}_1\times \mathcal{X}_2$ for any channel realization. Lemma \ref{Lemma Existence of fixed-point} is then an immediate result following from Proposition \ref{existence of fixed-point}.

\subsection{Proof of Lemma \ref{Lemma sufficient condition for full-rank case}}

Following the equivalence built in (\ref{Definition fixed-point}), we can instead derive sufficient conditions for the uniqueness of fixed-point in mapping $\boldsymbol{\Phi}$. To do so, we need to use the following Proposition from \cite{agarwal2001fixed}. 

\begin{prop}[Uniqueness of fixed-point]
Let $\mathcal{T}: \mathcal{X}\mapsto\mathcal{X}$ be a mapping defined on a finite-dimensional set $\mathcal{X}$. If $\mathcal{T}$ is a contraction with respect to some norm $\|\cdot\|$, that is, there exists some scalar $\alpha\in [0,1)$ such that
\begin{equation}
\|\mathcal{T}(x^{(1)})-\mathcal{T}(x^{(2)})\|\leq\alpha\|x^{(1)}-x^{(2)}\|,\forall x^{(1)},x^{(2)}\in\mathcal{X}.
\end{equation}
Then, there exists at most one fixed point of $\mathcal{T}$.
\label{Uniqueness of fixed-point}
\end{prop} 

To apply Proposition \ref{Uniqueness of fixed-point} for the mapping $\boldsymbol{\Phi}$ in (\ref{strategy mapping}), we need to define a suitable norm on the vector space $\mathbb{H}^{M}\times\mathbb{H}^{M}$ in which the mapping $\boldsymbol{\Phi}$ is defined. Inspired by the work in \cite{scutari2008competitive}, we introduce the weighted-maximum norm, which is defined as:
\begin{equation}
\|(\boldsymbol{X}_1,\boldsymbol{X}_2)\|_{F}^{\boldsymbol{w}}\triangleq\max\limits_{i\in\{1,2\}}\frac{\|\boldsymbol{X}_i\|_{F}}{w_i},
\label{w-m norm}
\end{equation}
where $\boldsymbol{X}_i\in\mathbb{H}^M$, $\|\cdot\|_{F}$ is the Frobenius norm and $\boldsymbol{w}=[w_1, w_2]>0$ is any positive vector. 

Next, we focus on the contraction property of the mapping $\boldsymbol{\Phi}$ with respect to the weighted-maximum norm $\|\cdot\|_{F}^{\boldsymbol{w}}$. 
That is to show the conditions for the existence of $\alpha\in[0,1)$, $\boldsymbol{w}>0$ such that
\begin{eqnarray}
&&\left\|\boldsymbol{\Phi}\left(\boldsymbol{Q}_1^{(1)},\boldsymbol{Q}_2^{(1)}\right) -\boldsymbol{\Phi}\left(\boldsymbol{Q}_1^{(2)},\boldsymbol{Q}_2^{(2)}\right) \right\|_{F}^{\boldsymbol{w}}\leq
\alpha\left\| 
\left(\boldsymbol{Q}_1^{(1)},\boldsymbol{Q}_2^{(1)}\right)-\left(\boldsymbol{Q}_1^{(2)},\boldsymbol{Q}_2^{(2)}\right)
\right\|_F^{\boldsymbol{w}},\;\;
\label{contractive property}
\\
&&\forall\; 
\left(\boldsymbol{Q}_1,\boldsymbol{Q}_2\right)\in\mathcal{X}_1\times\mathcal{X}_2.\nonumber
\end{eqnarray}

For the convenience of the notation, we introduce $e_i$ and $e_{\boldsymbol{\Phi}_i}$ defined as
\begin{equation}
e_i=\left\| 
\boldsymbol{Q}_i^{(1)}-\boldsymbol{Q}_i^{(2)}
\right\|_F,\;\;
e_{\boldsymbol{\Phi}_i}\triangleq\left\|
\mathcal{B}_i\left(\boldsymbol{Q}_j^{(1)}\right) 
-\mathcal{B}_i\left(\boldsymbol{Q}_j^{(2)}\right) 
\right\|_F,
\end{equation}
where $i,j\in\{1,2\}$ and $i\neq j$. 

Given two sets of strategies  $\left(\boldsymbol{Q}_1^{(1)},\boldsymbol{Q}_2^{(1)}\right),\left(\boldsymbol{Q}_1^{(2)},\boldsymbol{Q}_2^{(2)}\right)\in\mathcal{X}_1\times \mathcal{X}_2$, we can rewrite the left and right side of (\ref{contractive property}) in terms of the vector $\boldsymbol{e}=[e_1,e_2]^T$ and $\boldsymbol{e}_{\boldsymbol{\Phi}}=[e_{\boldsymbol{\Phi}_1},e_{\boldsymbol{\Phi}_2}]^T$ as
\begin{eqnarray}
&&\|\boldsymbol{e}\|_{\infty}^{\boldsymbol{w}}=\left\| 
\left(\boldsymbol{Q}_1^{(1)},\boldsymbol{Q}_2^{(1)}\right)-\left(\boldsymbol{Q}_1^{(2)},\boldsymbol{Q}_2^{(2)}\right)
\right\|_F^{\boldsymbol{w}},\;\;\\
&&\|\boldsymbol{e}_{\boldsymbol{\Phi}}\|_{\infty}^{\boldsymbol{w}}=\left\|\boldsymbol{\Phi}\left(\boldsymbol{Q}_1^{(1)},\boldsymbol{Q}_2^{(1)}\right) -\boldsymbol{\Phi}\left(\boldsymbol{Q}_1^{(2)},\boldsymbol{Q}_2^{(2)}\right) \right\|_{F}^{\boldsymbol{w}},
\end{eqnarray}
where the norm $\|\boldsymbol{x}\|_{\infty}^{\boldsymbol{w}}$ of a vector $\boldsymbol{x}\in \mathbb{C}^{K\times1}$ is defined as $\|\boldsymbol{x}\|_{\infty}^{\boldsymbol{w}}\triangleq\max\limits_{i\in\{1\cdots K\}}\frac{\|{x}_i\|_{F}}{w_i}$ with some $\boldsymbol{w}>0$.

The inequality in (\ref{contractive property}) can then be rewritten as
\begin{equation}
\|\boldsymbol{e}_{\boldsymbol{\Phi}}\|_{\infty}^{\boldsymbol{w}}
\leq
\alpha\|\boldsymbol{e}\|_{\infty}^{\boldsymbol{w}}.
\end{equation} 
 
We start to assume that the direct channels $\boldsymbol{H}_{ij}$ are full row-rank. We continue to show the expression of $\alpha$ by using the similar procedure in \cite{scutari2008competitive}. First, we have 
\begin{eqnarray}
e_{\boldsymbol{\Phi}_i}
&=&
\left\|
\mathcal{B}_i\left(\boldsymbol{Q}_j^{(1)}\right) 
-\mathcal{B}_i\left(\boldsymbol{Q}_j^{(2)}\right) 
\right\|_F \label{radius ineq. 1} \nonumber\\
&\leq&
\left\| 
\boldsymbol{X}_0(\boldsymbol{Q}_j^{(1)})-\boldsymbol{X}_0(\boldsymbol{Q}_j^{(2)})
\right\|_F \label{radius ineq. 2}\\
&=&
\left\| 
\left( \left(\eta_{ij}\boldsymbol{H}_{ij}^{\dag}\boldsymbol{\Sigma}_j^{(1)-1}\boldsymbol{H}_{ij}\right)^{-1}+ c_j\boldsymbol{P}_{\mathcal{N}(\boldsymbol{H}_{ij})}
\right) -
\left( 
\left(\eta_{ij}\boldsymbol{H}_{ij}^{\dag}\boldsymbol{\Sigma}_j^{(2)-1}\boldsymbol{H}_{ij}\right)^{-1}+
c_j\boldsymbol{P}_{\mathcal{N}(\boldsymbol{H}_{ij})}
\right) 
\right\|_F \IEEEeqnarraynumspace 
\label{radius ineq. 3}\\
&=&
\left\| 
\frac{\beta}{\gamma_{j}}\boldsymbol{H}_{ij}^{-1}
\boldsymbol{H}_{jj}\left( \text{diag}\left(\boldsymbol{Q}_j^{(1)}\right)-\text{diag}\left(\boldsymbol{Q}_j^{(2)}\right)
\right) 
\boldsymbol{H}_{jj}^{\dag} 
\boldsymbol{H}_{ij}^{-{\dag}}
\right\|_F \label{radius ineq. 4}\\
&\leq&
\rho\left(\frac{\beta}{\gamma_j}\boldsymbol{H}_{jj}^{\dag} 
\boldsymbol{H}_{ij}^{-{\dag}}\boldsymbol{H}_{ij}^{-1}\boldsymbol{H}_{jj}\right)\cdot 
\left\| 
 \text{diag}\left(\boldsymbol{Q}_j^{(1)}\right)-\text{diag}\left(\boldsymbol{Q}_j^{(2)}\right)
\right\|_F \label{radius ineq. 5}\\
&\leq&
\rho\left(\frac{\beta}{\gamma_{j}}\boldsymbol{H}_{jj}^{\dag} 
\boldsymbol{H}_{ij}^{-{\dag}}\boldsymbol{H}_{ij}^{-1}\boldsymbol{H}_{jj}\right)\cdot 
\left\| 
\boldsymbol{Q}_j^{(1)}-\boldsymbol{Q}_j^{(2)}
\right\|_F \nonumber\\
&=&
\rho\left(\frac{\beta}{\gamma_{j}}\boldsymbol{H}_{jj}^{\dag} 
\boldsymbol{H}_{ij}^{-{\dag}}\boldsymbol{H}_{ij}^{-1}\boldsymbol{H}_{jj}\right)e_j \nonumber
\end{eqnarray}
where (\ref{radius ineq. 2}) follows from inequality (\ref{non-expansive property}); (\ref{radius ineq. 3}) follows from (\ref{shortest distance problem});  
(\ref{radius ineq. 4}) follows from the reverse-order law for the generalized inverse i.e., $(\boldsymbol{A}^{\dag}\boldsymbol{X}\boldsymbol{A})^{-1}=\boldsymbol{A}^{-1}\boldsymbol{X}^{-1}\boldsymbol{A}^{-\dag}$ \cite{palomar2010convex}, which is valid if $\boldsymbol{A}$ is full row-rank matrix; (\ref{radius ineq. 5}) follows from the triangle inequality $\|\boldsymbol{AXA}^{\dag}\|_F\leq \rho(\boldsymbol{A}^{\dag}\boldsymbol{A})\|\boldsymbol{X}\|_F$.

Define the matrix $\boldsymbol{S}$ as follows:
\begin{equation}
{{[\boldsymbol{S}]}_{ij}}=
\left\lbrace 
\begin{split}
&
\rho\left(\frac{\beta}{\gamma_{j}}\boldsymbol{H}_{jj}^{\dag} 
\boldsymbol{H}_{ij}^{-{\dag}}\boldsymbol{H}_{ij}^{-1}\boldsymbol{H}_{jj}\right),\;\text{if}\; i\neq j,\\
& 0,\;\text{otherwise}.
\end{split}
\right. 
\label{judge matrix}
\end{equation}
 We then have
\begin{equation}
\left\| \boldsymbol{e}_{\Phi}\right\| ^{\boldsymbol{w}}_{\infty}
\leq 
\left\| \boldsymbol{S}\boldsymbol{e}\right\| ^{\boldsymbol{w}}_{\infty}
\leq 
\left\| \boldsymbol{S}\right\| ^{\boldsymbol{w}}_{\infty}\cdot\left\| \boldsymbol{e}\right\| ^{\boldsymbol{w}}_{\infty}.
\end{equation}
By setting $\alpha=\left\| \boldsymbol{S}\right\| ^{\boldsymbol{w}}_{\infty}$, $\boldsymbol{\Phi}$  in (\ref{strategy mapping}) is a contraction with respect to the weighted-maximum norm if there exists some $\boldsymbol{w}>0$ such that 
$\left\|\boldsymbol{S}\right\| ^{\boldsymbol{w}}_{\infty}<1.$
Note that all entries of $\boldsymbol{S}$ are non-negative. According to \cite{bertsekas1983distributed}, $\left\|\boldsymbol{S}\right\| ^{\boldsymbol{w}}_{\infty}<1$ for some $\boldsymbol{w}>0$ if and only if $\rho(\boldsymbol{S})<1$. Finally, the radius of $\boldsymbol{S}$ can be easily obtained
\begin{equation}
\rho(\boldsymbol{S})=\sqrt{\rho\left(\frac{\beta}{\gamma_1}\boldsymbol{H}_{11}^{\dag} 
\boldsymbol{H}_{21}^{-{\dag}}\boldsymbol{H}_{21}^{-1}\boldsymbol{H}_{11}\right)
\rho\left(\frac{\beta}{\gamma_2}\boldsymbol{H}_{22}^{\dag} 
\boldsymbol{H}_{12}^{-{\dag}}\boldsymbol{H}_{12}^{-1}\boldsymbol{H}_{22}\right)}\;,
\end{equation}
which leads to the condition (\ref{contractive condition}).

\subsection{Proof of Theorem \ref{Theorem conditions of unique NE}}
The main challenge in extending Lemma 4 to the general full-duplex channel is that the reverse-order law for the generalized inverse, which is used in (\ref{radius ineq. 4}), is valid only under the full row-rank assumption of $\boldsymbol{H}_{ij}$. In general, one can only have the inequality $(\boldsymbol{A}^{\dag}\boldsymbol{X}\boldsymbol{A})^{-1}\preceq\boldsymbol{A}^{-1}\boldsymbol{X}^{-1}\boldsymbol{A}^{-\dag}$. For the sake of simplicity, we first focus on the strictly full-rank case, where the direct channel matrices $\{\boldsymbol{H}_{ij}\}_{i,j\in\{1,2\}, i\neq j}$ are with full column-rank. Using the notations and following the procedure in the proof of Lemma 4, we obtain
\begin{eqnarray}
e_{\boldsymbol{\Phi}_i}
&=&
\left\|
\mathcal{B}_i\left(\boldsymbol{Q}_j^{(1)}\right) 
-\mathcal{B}_i\left(\boldsymbol{Q}_j^{(2)}\right) 
\right\|_F \label{gradius ineq. 1} \nonumber\\
&\leq&
\left\| 
\boldsymbol{X}_0(\boldsymbol{Q}_j^{(1)})-\boldsymbol{X}_0(\boldsymbol{Q}_j^{(2)})
\right\|_F \label{gradius ineq. 2}\nonumber\\
&=&
\left\| 
\left( \left(\eta_{ij}\boldsymbol{H}_{ij}^{\dag}\boldsymbol{\Sigma}_j^{(1)-1}\boldsymbol{H}_{ij}\right)^{-1}+ c_j\boldsymbol{P}_{\mathcal{N}(\boldsymbol{H}_{ij})}
\right) -
\left( 
\left(\eta_{ij}\boldsymbol{H}_{ij}^{\dag}\boldsymbol{\Sigma}_j^{(2)-1}\boldsymbol{H}_{ij}\right)^{-1}+
c_j\boldsymbol{P}_{\mathcal{N}(\boldsymbol{H}_{ij})}
\right) 
\right\|_F \IEEEeqnarraynumspace 
\label{gradius ineq. 3}\nonumber\\
&=&
\left\| 
 \left(\eta_{ij}\boldsymbol{H}_{ij}^{\dag}\boldsymbol{\Sigma}_j^{(1)-1}\boldsymbol{H}_{ij}\right)^{-1} 
 -
\left(\eta_{ij}\boldsymbol{H}_{ij}^{\dag}\boldsymbol{\Sigma}_j^{(2)-1}\boldsymbol{H}_{ij}\right)^{-1} 
\right\|_F
 \label{gradius ineq. 4}
\end{eqnarray}
The strictly full column-rank of $\boldsymbol{H}_{ij}$ implies that $\boldsymbol{H}_{ij}$ is deficient in row-rank, thus we cannot derive the upper-bound for (\ref{gradius ineq. 4}), as was done in (\ref{radius ineq. 4}). Thus, we need to develop an approach suitable for deriving the upper-bound in the full column-rank case. To this end, we introduce the function $\boldsymbol{P}_i(\boldsymbol{X}): \mathcal{Q}_i \mapsto\mathbb{C}^{M\times M}$ as follows,
\begin{equation}
\boldsymbol{P}_i(\boldsymbol{X})=\left(\boldsymbol{H}_{ij}^{\dag}\left( \boldsymbol{I}+\beta\eta_{jj}\boldsymbol{H}_{jj}\boldsymbol{X}\boldsymbol{H}_{jj}^{\dag}\right) ^{-1}\boldsymbol{H}_{ij}\right)^{-1},
\label{PX def}
\end{equation} 
where $\mathcal{Q}_i\triangleq\{\boldsymbol{Q}\in\mathbb{H}_{+}^M|\textbf{tr}(\boldsymbol{Q})= P_i\}$ is a convex set. Since $\boldsymbol{H}_{ij}$ is with the full column-rank that $\boldsymbol{P}_i(\boldsymbol{X})$ is invertible, which implies that $\boldsymbol{P}_i(\boldsymbol{X})$ is continuous on $\mathcal{Q}_i$ and differentiable on the interior of $\mathcal{Q}_i$ \cite{rao1985generalized}. 
Invoking the mean-value theorem in \cite{palomar2010convex}, 
we obtain that for any given $\boldsymbol{X},\boldsymbol{Y}\in \mathcal{Q}_i$ there exists  $\boldsymbol{\xi}=\alpha\boldsymbol{X}+\beta\boldsymbol{Y}$ with some $\alpha, \beta\geq 0$ and $\alpha+\beta=1$ such that
\begin{equation}
\|\boldsymbol{P}_i(\boldsymbol{Y})-\boldsymbol{P}_i(\boldsymbol{X})\|_F\leq\|\boldsymbol{D_XP}_i(\boldsymbol{\xi})\|_2\|\boldsymbol{Y}-\boldsymbol{X}\|_F,
\label{mean-value theorem}
\end{equation}
where $\boldsymbol{D_XP}_i$ denotes the derivative of the function $\boldsymbol{P}_i(\boldsymbol{X})$ with respect to $\boldsymbol{X}$, defined by the following equation \cite{hjorungnes2007complex}: 
\begin{equation}
\text{dvec}(\boldsymbol{P}_i)=(\boldsymbol{D_XP}_i)\text{dvec}(\boldsymbol{X})+(\boldsymbol{D}_{\boldsymbol{X}^{\dag}}\boldsymbol{P}_i)\text{dvec}(\boldsymbol{X}^{\dag}).
\label{identity rule}
\end{equation}
The above identity implies that we can derive $\boldsymbol{D_XP}_i$ by differentiating and vectorizing $\boldsymbol{P}_i(\boldsymbol{X})$. For the convenience of notation, we denote $\boldsymbol{R}_j(\boldsymbol{X})=\boldsymbol{I}+\beta\eta_{jj}\boldsymbol{H}_{jj}\boldsymbol{X}\boldsymbol{H}_{jj}^{\dag}$, simplified as $\boldsymbol{R}_j$. Then,
\begin{eqnarray}
\text{dvec}(\boldsymbol{P}_i)&=&\text{vec}(\text{d}\boldsymbol{P}_i)=-\text{vec}(\boldsymbol{P}_i(\text{d}\boldsymbol{P}_i^{-1})\boldsymbol{P}_i)\nonumber\\
&=&\text{vec}\left(\boldsymbol{P}_i\cdot\text{d}\left(\boldsymbol{H}_{ij}^{\dag}\boldsymbol{R}_j^{-1}\boldsymbol{H}_{ij}\right)\cdot\boldsymbol{P}_i\right)\label{dp2}\\
&=&\beta\eta_{jj}\text{vec}\left(\boldsymbol{P}_i\boldsymbol{H}_{ij}^{\dag}\boldsymbol{R}_j^{-1}\boldsymbol{H}_{jj}\cdot\text{d}\boldsymbol{(X)}\cdot\boldsymbol{H}_{jj}^{\dag}\boldsymbol{R}_j^{-1}\boldsymbol{H}_{ij}\boldsymbol{P}_i\right)\nonumber\\
&=&\beta\eta_{jj}\text{vec}\left(\boldsymbol{G}_i(\boldsymbol{X})\text{d}\boldsymbol{(X)}\overline{\boldsymbol{G}_i(\boldsymbol{X})}\right)\nonumber\\
&=&\beta\eta_{jj}\left(\overline{\boldsymbol{G}_i(\boldsymbol{X})}\otimes\boldsymbol{G}_i(\boldsymbol{X})\right)\text{dvec}(\boldsymbol{X}),\label{dp5}
\end{eqnarray}
where $\boldsymbol{G}_i(\boldsymbol{X})=\boldsymbol{P}_i\boldsymbol{H}_{ij}^{\dag}\boldsymbol{R}_j^{-1}\boldsymbol{H}_{jj}$ is a Hermitian matrix; $\otimes$ denotes the Kronecker product; (\ref{dp2}) follows from the inversibility of $\boldsymbol{P}_i(\boldsymbol{X})$; (\ref{dp5}) follows from the result in \cite{horn2012matrix}. (\ref{identity rule}) and (\ref{dp5}) indicate that $\boldsymbol{D_XP}_i=\beta\left(\overline{\boldsymbol{G}_i(\boldsymbol{X})}\otimes\boldsymbol{G}_i(\boldsymbol{X})\right)$. Invoking (\ref{mean-value theorem}), we obtain
\begin{equation}
\|\boldsymbol{P}_i(\boldsymbol{Y})-\boldsymbol{P}_i(\boldsymbol{X})\|_F
\leq
\beta\eta_{jj}\|\overline{\boldsymbol{G}_i(\boldsymbol{\xi})}\otimes\boldsymbol{G}_i(\boldsymbol{\xi})\|_2\|\boldsymbol{Y}-\boldsymbol{X}\|_F.
\label{alternative mean-value theorem}
\end{equation} 
We can further investigate (\ref{dp5}) by
\begin{eqnarray}
\|\overline{\boldsymbol{G}_i(\boldsymbol{\xi})}\otimes\boldsymbol{G}_i(\boldsymbol{\xi})\|_2&=&
\sqrt{\rho\left( 
\left( \overline{\boldsymbol{G}_i(\boldsymbol{\xi})}\otimes\boldsymbol{G}_i(\boldsymbol{\xi})\right)^{\dag} 
\left( \overline{\boldsymbol{G}_i(\boldsymbol{\xi})}\otimes\boldsymbol{G}_i(\boldsymbol{\xi})\right) \right)}\nonumber\\
&=&\sqrt{\rho\left( 
\left( {\boldsymbol{G}_i(\boldsymbol{\xi})}^T\otimes\boldsymbol{G}_i^{\dag}(\boldsymbol{\xi})\right) 
\left( \overline{\boldsymbol{G}_i(\boldsymbol{\xi})}\otimes\boldsymbol{G}_i(\boldsymbol{\xi}) \right)
\right)}\nonumber\\
&=&\sqrt{\rho\left( 
\left( \boldsymbol{G}_i^T(\boldsymbol{\xi})\overline{\boldsymbol{G}_(\boldsymbol{\xi})}\right) 
\otimes
\left( \boldsymbol{G}_i^{\dag}(\boldsymbol{\xi})\boldsymbol{G}_i(\boldsymbol{\xi})\right) \right)}\nonumber\\
&=&\rho 
\left( \boldsymbol{G}_i^{\dag}(\boldsymbol{\xi})\boldsymbol{G}_i(\boldsymbol{\xi})\right)
\label{G function}.
\end{eqnarray}
Each step in (\ref{G function}) follows from the property of the Kronecker product. Note that the orthogonal projection onto the range of $\boldsymbol{R}_j^{-1/2}\boldsymbol{H}_{ij}^{\dag}$ i.e., $\text{Ran}(\boldsymbol{R}_j^{-1/2}\boldsymbol{H}_{ij}^{\dag})$, is defined as 
\begin{equation}
\boldsymbol{P}_{\boldsymbol{R}_j}=\boldsymbol{R}_j^{-1/2}\boldsymbol{H}_{ij}\boldsymbol{P}_i\boldsymbol{H}_{ij}^{\dag}\boldsymbol{R}_j^{-1/2}.
\end{equation}
Then, $\boldsymbol{G}_i(\boldsymbol{\xi})$ can be rewritten in terms of $\boldsymbol{P}_{\boldsymbol{R}_j}$ as $\boldsymbol{G}_i(\boldsymbol{\xi})=\boldsymbol{H}_{ij}^{-1}\boldsymbol{R}_j^{1/2}\boldsymbol{P}_{\boldsymbol{R}_j}\boldsymbol{R}_j^{-1/2}\boldsymbol{H}_{jj}$. It follows that
\begin{eqnarray}
\rho 
\left( \boldsymbol{G}_i^{\dag}(\boldsymbol{\xi})\boldsymbol{G}_i(\boldsymbol{\xi})\right)
&=&\rho
\left(
\boldsymbol{H}_{jj}^{\dag}\boldsymbol{R}_j^{-1/2}\boldsymbol{P}_{\boldsymbol{R}_j}^{\dag}\boldsymbol{R}_j^{1/2}\boldsymbol{H}_{ij}^{-\dag}
\boldsymbol{H}_{ij}^{-1}\boldsymbol{R}_j^{1/2}\boldsymbol{P}_{\boldsymbol{R}_j}\boldsymbol{R}_j^{-1/2}\boldsymbol{H}_{jj}
\right)\nonumber\\
&\leq&
\rho
\left(
\boldsymbol{H}_{jj}^{\dag}\boldsymbol{H}_{jj}
\right)
\rho
\left(\boldsymbol{R}_j^{-1}(\boldsymbol{\xi})\right)
\rho
\left(\boldsymbol{H}_{ij}^{-\dag}\boldsymbol{H}_{ij}^{-1}
\right)
\rho
\left(\boldsymbol{R}_j(\boldsymbol{\xi})
\right).
\label{bound on G}
\end{eqnarray}
Let $\boldsymbol{X}=\text{diag}(\boldsymbol{Q}_j^{(1)}), \boldsymbol{Y}=\text{diag}(\boldsymbol{Q}_j^{(2)})$. We can continue estimate (\ref{gradius ineq. 4}),
\begin{eqnarray}
e_{\boldsymbol{\Phi}_i}
&\leq&
\left\| 
 \left(\eta_{ij}\boldsymbol{H}_{ij}^{\dag}\boldsymbol{\Sigma}_j^{(1)-1}\boldsymbol{H}_{ij}\right)^{-1} 
 -
\left(\eta_{ij}\boldsymbol{H}_{ij}^{\dag}\boldsymbol{\Sigma}_j^{(2)-1}\boldsymbol{H}_{ij}\right)^{-1} 
\right\|_F\nonumber\\
&=&
\frac{1}{\eta_{ij}}
\|\boldsymbol{P}_i(\boldsymbol{Y})-\boldsymbol{P}_i(\boldsymbol{X})\|_F\label{bound of Dx 1}\\
&\leq&
\frac{\beta\eta_{jj}}{\eta_{ij}}
\rho
\left(
\boldsymbol{H}_{jj}^{\dag}\boldsymbol{H}_{jj}
\right)
\rho
\left(\boldsymbol{H}_{ij}^{-\dag}\boldsymbol{H}_{ij}^{-1}
\right)
\rho
\left(\boldsymbol{R}_j^{-1}(\boldsymbol{\xi})\right)
\rho
\left(\boldsymbol{R}_j(\boldsymbol{\xi})
\right)\|\boldsymbol{Y}-\boldsymbol{X}\|_F
\label{bound of Dx 2}\\
&\leq&
\frac{\beta\eta_{jj}}{\eta_{ij}}
\left(1+\beta \eta _{jj}P_j\rho\left(\boldsymbol{H}_{jj}^{\dag}\boldsymbol{H}_{jj}
\right)\right)
\rho
\left(
\boldsymbol{H}_{jj}^{\dag}\boldsymbol{H}_{jj}
\right)
\rho
\left(\boldsymbol{H}_{ij}^{-\dag}\boldsymbol{H}_{ij}^{-1}
\right)\|\boldsymbol{Q}_j^{(1)}-\boldsymbol{Q}_j^{(2)}\|_F
\label{bound of Dx 3}\\
&=&
\frac{\beta\eta_{jj}}{\eta_{ij}}
\left(1+\beta \eta _{jj}P_j\rho\left(\boldsymbol{H}_{jj}^{\dag}\boldsymbol{H}_{jj}
\right)\right)
\rho
\left(
\boldsymbol{H}_{jj}^{\dag}\boldsymbol{H}_{jj}
\right)
\rho
\left(\boldsymbol{H}_{ij}^{-\dag}\boldsymbol{H}_{ij}^{-1}
\right)e_j.
\label{bound of Dx 4}
\end{eqnarray}
where (\ref{bound of Dx 1}) follows from the definition of $\boldsymbol{P}_i(\cdot)$ in (\ref{PX def}); (\ref{bound of Dx 2}) follows from (\ref{alternative mean-value theorem}), (\ref{G function}) and (\ref{bound on G}). To obtain (\ref{bound of Dx 3}), we use the inequality
\begin{eqnarray}
\boldsymbol{I}&\preceq&\boldsymbol{R}_j(\boldsymbol{\xi})
\preceq \boldsymbol{I}+\beta \eta_{jj}P_j\boldsymbol{H}_{jj}^{\dag}\boldsymbol{H}_{jj}.
\end{eqnarray}
By using (\ref{bound of Dx 4}), we can then derive the sufficient condition of the unique NE for the full column-rank case. The derivation is same as in Lemma \ref{Lemma sufficient condition for full-rank case}.

Next, we extend the sufficient condition to the rank deficient case. In order to use the results for the full-rank case, the rank-deficient matrix $\boldsymbol{H}_{ij}$ should be modified to a lower-dimensional full-rank matrix $\widetilde{\boldsymbol{H}}_{ij}$, where $\widetilde{\boldsymbol{H}}_{ij}\in \mathbb{C}^{N\times r}$ with $r=\text{rank}(\boldsymbol{H}_{ij})$. The singular value decomposition (SVD) writes $\boldsymbol{H}_{ij}$ as $\boldsymbol{H}_{ij}=\boldsymbol{U}_{ij}\boldsymbol{\Sigma}_{ij}\boldsymbol{V}^{\dag}_{ij}$ where $\boldsymbol{U}_{ij}$ and $\boldsymbol{V}_{ij}$ are semi-unitary matrices; $\boldsymbol{V}_{ij}$ is a basis for the row space of $\boldsymbol{H}_{ij}$.
Define $\widetilde{\boldsymbol{H}}_{ij}=\boldsymbol{H}_{ij}\boldsymbol{V}_{ij}$ and replace $\boldsymbol{H}_{ij}$ by $\widetilde{\boldsymbol{H}}_{ij}$. Then, the optimization problem in (\ref{NE game 1}) is modified to the following lower-dimensional optimization problem:
 \begin{equation}
   \begin{split}
   \max\limits_{\widetilde{\boldsymbol{Q}_i}}\;\;& \text{log}\;\text{det}(\boldsymbol{I}+\eta_{ij}\widetilde{\boldsymbol{H}}_{ij}^{\dag}\boldsymbol{\Sigma}_{j}^{-1}\widetilde{\boldsymbol{H}}_{ij}\widetilde{\boldsymbol{Q}_i})\\
   \text{subject to}\;\;
   & \widetilde{\boldsymbol{Q}_i}\in\widetilde{\mathcal{X}_i},
   \end{split} 
   \label{modified NE game}
  \end{equation}  
where $\widetilde{\mathcal{X}_i}=\left\lbrace\widetilde{\boldsymbol{Q}_i}\in\mathbb{H}^{r}|\widetilde{\boldsymbol{Q}_i}\succeq0,\textbf{tr}\left(\widetilde{\boldsymbol{Q}_i}\right)\leq P_i\right\rbrace$ and $\widetilde{\boldsymbol{H}}_{ij}$ is full column-rank matrix. Denote the optimal solution of problem (\ref{modified NE game}) as $\mathcal{B}_{r_i}(\boldsymbol{Q}_j)$. Next, by building the equivalence between problem (\ref{original waterfilling problem}) and problem (\ref{modified NE game}) we can show that $\mathcal{B}_i(\boldsymbol{Q}_j)=\boldsymbol{V}_{ij}\mathcal{B}_{r_i}(\boldsymbol{Q}_j)\boldsymbol{V}_{ij}^{\dag}$. 

First, for any matrix $\boldsymbol{Q}_i\in\mathcal{X}_i$, there exists some $\widetilde{\boldsymbol{Q}_i}\in\widetilde{\mathcal{X}_i}$ such that $\boldsymbol{Q}_i=\boldsymbol{V}_{ij}\widetilde{\boldsymbol{Q}_i}\boldsymbol{V}_{ij}^{\dag}$. Thus, $\mathcal{B}_i(\boldsymbol{Q}_i)=\boldsymbol{V}_{ij}\widetilde{\boldsymbol{Q}_i}^*\boldsymbol{V}_{ij}^{\dag}$, $\widetilde{\boldsymbol{Q}_i}^*\in\widetilde{\mathcal{X}}_i$. Then, using the identity $\text{det}(\boldsymbol{I}+\boldsymbol{AB})=\text{det}(\boldsymbol{I}+\boldsymbol{BA})$ we obtain
\begin{eqnarray}
&&\text{det}(\boldsymbol{I}+\eta_{ij}{\boldsymbol{H}}_{ij}^{\dag}\boldsymbol{\Sigma}_{j}^{-1}{\boldsymbol{H}}_{ij}\mathcal{B}_i(\boldsymbol{Q}_j)) \nonumber\\
&=&\text{det}(\boldsymbol{I}+\eta_{ij}{\boldsymbol{H}}_{ij}\mathcal{B}_i(\boldsymbol{Q}_i){\boldsymbol{H}}_{ij}^{\dag}\boldsymbol{\Sigma}_{i}^{-1}) \nonumber\\
&=&\text{det}(\boldsymbol{I}+\eta_{ij}{\boldsymbol{H}}_{ij}\boldsymbol{V}_{ij}\widetilde{\boldsymbol{Q}_i}^*\boldsymbol{V}_{ij}^{\dag}{\boldsymbol{H}}_{ij}^{\dag}\boldsymbol{\Sigma}_{i}^{-1}) \nonumber\\
&=&\text{det}(\boldsymbol{I}+\eta_{ij}{\widetilde{\boldsymbol{H}}_{ij}}\widetilde{\boldsymbol{Q}_i}^*\widetilde{{\boldsymbol{H}}}_{ij}^{\dag}\boldsymbol{\Sigma}_{i}^{-1}) \nonumber\\
&=&\text{det}(\boldsymbol{I}+\eta_{ij}{\widetilde{{\boldsymbol{H}}}_{ij}^{\dag}\boldsymbol{\Sigma}_{i}^{-1}\widetilde{\boldsymbol{H}}_{ij}}\widetilde{\boldsymbol{Q}_i}^*). \nonumber
\end{eqnarray}
$\widetilde{\boldsymbol{Q}_i}^*$ is a feasible solution of problem (\ref{modified NE game}), implying that the maximum of problem (\ref{modified NE game}) is no smaller than that of problem (\ref{original waterfilling problem}). On the other hand, $\boldsymbol{V}_{ij}\mathcal{B}_{r_i}(\boldsymbol{Q}_j)\boldsymbol{V}_{ij}^{\dag}$ is feasible for problem (\ref{original waterfilling problem}). Thus the maximum of problem (\ref{modified NE game}) is no greater than that of problem (\ref{original waterfilling problem}). Therefore, problem (\ref{original waterfilling problem}) and problem (\ref{modified NE game}) are equivalent, implying $\mathcal{B}_i(\boldsymbol{Q}_j)=\boldsymbol{V}_{ij}\mathcal{B}_{r_i}(\boldsymbol{Q}_j)\boldsymbol{V}_{ij}^{\dag}$.

We now can continue to prove Theorem \ref{Theorem conditions of unique NE}. Similar to the proof of the full-rank case, we have
\begin{eqnarray}
e_{\boldsymbol{\Phi}_i}
&=&
\left\|
\mathcal{B}_i\left(\boldsymbol{Q}_j^{(1)}\right) 
-\mathcal{B}_i\left(\boldsymbol{Q}_j^{(2)}\right)
\right\|_F \nonumber\\
&=&
\left\|
\boldsymbol{V}_{ij}\mathcal{B}_{r_i}\left(\boldsymbol{Q}_j^{(1)}\right)\boldsymbol{V}_{ij}^{\dag} 
-\boldsymbol{V}_{ij}\mathcal{B}_{r_i}\left(\boldsymbol{Q}_j^{(2)}\right)\boldsymbol{V}_{ij}^{\dag} 
\right\|_F  \nonumber\\
&\leq&
\rho(\boldsymbol{V}_{ij}^{\dag}\boldsymbol{V}_{ij})\left\|
\mathcal{B}_{r_i}\left(\boldsymbol{Q}_j^{(1)}\right) 
-\mathcal{B}_{r_i}\left(\boldsymbol{Q}_j^{(2)}\right)
\right\|_F \label{general radius 2}\\
&=&
\left\|
\mathcal{B}_{r_i}\left(\boldsymbol{Q}_j^{(1)}\right) 
-\mathcal{B}_{r_i}\left(\boldsymbol{Q}_j^{(2)}\right)
\right\|_F \nonumber\\
&\leq&
\frac{\beta}{\gamma_{j}}
\left(1+\beta \eta _{jj}P_j\rho\left(\boldsymbol{H}_{jj}^{\dag}\boldsymbol{H}_{jj}
\right)\right)
\rho
\left(
\boldsymbol{H}_{jj}^{\dag}\boldsymbol{H}_{jj}
\right)
\rho
\left(\widetilde{\boldsymbol{H}}_{ij}^{-\dag}\widetilde{\boldsymbol{H}}_{ij}^{-1}
\right)
\left\| 
\boldsymbol{Q}_j^{(1)}-\boldsymbol{Q}_j^{(2)}
\right\|_F \label{general radius 3}\\
&=&
\frac{\beta}{\gamma_{j}}
\left(1+\beta \eta _{jj}P_j\rho\left(\boldsymbol{H}_{jj}^{\dag}\boldsymbol{H}_{jj}
\right)\right)
\rho
\left(
\boldsymbol{H}_{jj}^{\dag}\boldsymbol{H}_{jj}
\right)
\rho
\left(\widetilde{\boldsymbol{H}}_{ij}^{-\dag}\widetilde{\boldsymbol{H}}_{ij}^{-1}
\right)e_j \nonumber
\label{generalized judge inequality}
\end{eqnarray}
where (\ref{general radius 2}) follows from the the triangle inequality $\|\boldsymbol{AXA}^{\dag}\|_F\leq \rho(\boldsymbol{A}^{\dag}\boldsymbol{A})\|\boldsymbol{X}\|_F$; (\ref{general radius 3}) follows from (\ref{bound of Dx 4}). 
The relationship between $\boldsymbol{H}_{ij}$ and $\widetilde{\boldsymbol{H}}_{ij}$ implies that
\begin{equation}
\rho
\left(\widetilde{\boldsymbol{H}}_{ij}^{-\dag}\widetilde{\boldsymbol{H}}_{ij}^{-1}
\right)=\rho
\left({\boldsymbol{H}}_{ij}^{-\dag}{\boldsymbol{H}}_{ij}^{-1}
\right).
\end{equation}
The rest of the proof can then be completed following the same steps in Lemma 4.


%





\ifCLASSOPTIONcaptionsoff
  \newpage
\fi

\small
\bibliographystyle{IEEEtran}
\bibliography{IEEEabrv,myMSbib}
\end{document}